\documentclass[showkeys,12pt,onecolumn]{article}

\usepackage{arxiv}
\usepackage[english]{babel}
\usepackage[utf8]{inputenc}
\usepackage[colorinlistoftodos, color=green!40, prependcaption]{todonotes}
\usepackage[colorlinks=true,linkcolor=blue,urlcolor=blue,citecolor=blue]{hyperref}
\usepackage{mathrsfs}
\usepackage{amsmath}
\usepackage{amsfonts}
\usepackage{amssymb}
\usepackage{amsthm}
\usepackage{mathdesign}
\usepackage{slashed}
\usepackage{mathtools}
\usepackage{xcolor}
\usepackage{enumitem}
\usepackage[capitalise]{cleveref}
\usepackage[normalem]{ulem}
\usepackage{float}
\usepackage{cancel}
\usepackage[sorting = none, backend = bibtex, style=numeric-comp]{biblatex}

\makeatletter
\newcommand{\mybf}[1]{%
    \begingroup
    \def\@tempa{#1}%
    \ifx\@tempa\@empty
        \textbf{#1}%
    \else
        \edef\@tempa{\noexpand\in@{#1}{abcdefghijklmnopqrstuvwxyzABCDEFGHIJKLMNOPQRSTUVWXYZ}}%
        \@tempa
        \ifin@
            \mathbf{#1}%
        \else
            \pmb{#1}%
        \fi
    \fi
    \endgroup
}
\makeatother

\begin{document}

\title{Lattice Boltzmann method for warm fluid simulations of plasma wakefield acceleration}

\author{Daniele Simeoni$^{1,*}$
\And Gianmarco Parise$^{2,3}$
\And Fabio Guglietta$^{1}$
\And Andrea Renato Rossi$^{4}$
\And James Rosenzweig$^{5}$
\And Alessandro Cianchi$^{1}$
\And Mauro Sbragaglia$^{1}$ %
\and \\
$^{1}$Department of Physics \& INFN, Tor Vergata University of Rome, Via della Ricerca Scientifica 1, 00133, Rome, Italy
\and
$^{2}$Department of Physics, Tor Vergata University of Rome, Via della Ricerca Scientifica 1, 00133, Rome, Italy
\and
$^{3}$INFN, Laboratori Nazionali di Frascati, Via Enrico Fermi 54, 00044, Frascati, Italy
\and
$^{4}$INFN, Section of Milan, via Celoria 16, 20133, Milan, Italy
\and
$^{5}$Department of Physics and Astronomy, University of California-Los Angeles,\\ Portola Plaza 475, 90095, Los Angeles, CA, USA
\and
$^{*}$\texttt{daniele.simeoni@roma2.infn.it}
}

\date{\today} 

\maketitle


\begin{abstract}
A comprehensive characterization of lattice Boltzmann (LB) schemes to perform warm fluid numerical simulations of particle wakefield  acceleration (PWFA) processes is discussed in this paper. The LB schemes we develop hinge on the {\it moment matching} procedure, allowing the fluid description of a {\it warm relativistic} plasma wake generated by a driver pulse propagating in a neutral plasma. We focus on fluid models equations resulting from two popular closure assumptions of the relativistic kinetic equations, i.e., the {\it local equilibrium} and the {\it warm plasma} closure assumptions. The developed LB schemes can thus be used to disclose insights on the quantitative differences between the two closure approaches in the dynamics of PWFA processes. Comparisons between the proposed schemes and available analytical results are extensively addressed.
\end{abstract}

\section{Introduction} \label{sec:intro}

The process of particle acceleration plays a role of primary importance at the interface between fundamental and applied physics~\cite{gourlay-2022}. The growing costs and dimensions of conventional large-scale accelerators demand for new and more efficient technologies to push the energies reached by particle beams beyond the state of the art of modern day capabilities. In this context, plasma acceleration is a promising technique that would enable the construction of compact particle accelerators, while retaining the same (or superior) energy gains obtained with conventional methods~\cite{faure-2004, blumenfeld-2007, litos-2014}. A ionized gas is perturbed via the injection of relativistic charged particles (particle wakefield acceleration, PWFA)~\cite{chen-1985} or intense lasers (laser wakefield acceleration, LWFA)~\cite{tajima-1979}, generally named \emph{driver}: the interaction between the neutral plasma and the injected driver creates a wave like dynamics of positive and negative charges and hence strong accelerating fields (up to $100$ GV/m) are developed; the interested reader might look into~\cite{bingham-2004,esarey-2009} to go into detail on the topic. The described process involves a large number of "actors": the injected driver components, whether particles moving near the speed of light – typically electrons – or laser fields, and both the ions and electrons that make up the plasma. All of them interact with each other via electromagnetic forces, thus making it really difficult to predict and control the final behavior of a plasma acceleration experiment. Theoretical modelling and numerical simulations are therefore a powerful tool to help guide the design of new experiments. \\
On the side of numerical simulations, the most commonly used techniques in the field are represented by particle in cell (PIC) methods~\cite{birdsall-2018, fonseca-2002, lehe-2016, burau-2010, benedetti-2008}, that employ single particle dynamics to describe both the particles in the driver (in this paper we will focus on PWFA) and the plasma components. These schemes are deeply fine-grained and, while they can capture the most refined phenomena that happen at the microscopic scale, they bring in low-statistics numerical noise~\cite{birdsall-2018, hockney-2021}. An alternative numerical modelling can be proposed by recurring to continuum fluid descriptions of the system. Fluid models describe the plasma via macroscopic fields such as particle number density and fluid velocity, and they do so by solving the inviscid relativistic Euler equations that can be systematically derived from kinetic theory of charged gases via a coarse graining of the relativistic Maxwell-Vlasov system~\cite{cercignani-2002, degroot-1980, rezzolla-2013} (in their most general warm formulation these equations are not closed and hence require additional constrains -- more on this later on). Despite losing the ability to describe some kinetically pertinent features, numerical methods relying on the fluid description are still able to capture non trivial features of the PWFA system, and are set up by construction not to show statistical noise. Some examples of fluid solvers used in the realm of PWFA are Architect~\cite{marocchino-2016}, QFLUID~\cite{tomassini-2015}, MARPLE~\cite{boldarev-2012}, FLASH~\cite{cook-2016} and the code used for hydrodynamic optically-field-ionized plasma channels in~\cite{mewes-2023}. \\
From the theoretical point of view, fluid models have traditionally been developed by neglecting thermal effects, i.e., by neglecting pressure terms in the Euler equations. This choice has been motivated first and foremost by the fact that initial electron thermal energy in the plasma is expected to be of the order of $k_b T_{i} \sim 10$ eV~\cite{anania-2014, gonsalves-2019}, which is a small value when compared with the electron rest energy $m_e c^2 = 0.511$ MeV (the initial thermal energy normalized to the electron rest mass, $\mu_i = k_b T_i / (m_e c^2)$, is the control parameter that is usually used to assess the importance of thermal effects), and secondly by the fact that these cold fluid models are easier to treat theoretically, providing even analytical results in some simplified cases~\cite{ruth-1984, chen-1987, lu-2005}. \\
Nevertheless, there is a series of important reasons that drives the development of warm fluid models for PWFA. First, the wave-like solutions to cold fluid equations become singular in the presence of wide and highly charged driver pulses (\textit{Wave Breaking}~\cite{akhiezer-1956, dawson-1959}). The presence of thermal effects may be one of the regularising mechanisms that mitigate the singularity~\cite{katsouleas-1988, rosenzweig-1988, schroeder-2005, jain-2015}, by altering the wakefield properties and allowing electron trapping in the wakefield~\cite{esarey-2007}. A significant heating is also expected in the post-wavebreaking dynamic~\cite{lotov-2003}. Second, studies targeted at the late stage dynamics of the process~\cite{gholizadeh-2011, gilljohann-2019, darcy-2022, zgadzaj-2020,khudiakov-2022} point to the importance of electrons temperature (with particular emphasis on thermally driven ion-acoustic motion~\cite{silva-2021,sahai-2014,sahai-2017,khudiakov-2022,zgadzaj-2020}) for the restoration of the equilibrium conditions after a driver pulse has passed in the plasma channel. Understanding the restoration conditions is pivotal to enable the possibility of having high repetition rates of driver pulses in order to create sustained accelerating fields. Third, although the aforementioned initial temperatures would not lead to meaningful divergences in behavior from the cold case (at least in the first wave periods~\cite{lotov-2003, esarey-2007}), a different situation is expected as many consecutive pulses are injected into the system. The energy deposited by every pulse would be partially transferred to the plasma~\cite{gholizadeh-2011, darcy-2022}, leading to significant increases in temperature (some estimates provide $O(1)$ keV increases in post wave-breaking situations~\cite{zgadzaj-2020, khudiakov-2022}). Lastly, we mention that temperature effects might be also relevant for the study of positron acceleration in quasi-hollow warm plasma channels ~\cite{silva-2021,wang-2021,diederichs-2023,diederichs-2023-b}. \\
If one wants to develop a warm fluid theory, often the only viable option when trying to derive analytical results, an immediate problem has to be tackled: additional fields are now present in the set of equations (namely the pressure tensor fields) and therefore a suitable closure has to be carefully selected. The closure problem has been studied in the community, and various models have been proposed~\cite{toepfer-1971, newcomb-1982, amendt-1985, newcomb-1986-b, siambis-1987, pennisi-1991} (see~\cite{muscato-1993} for an outlook).\\
In this paper we explore two closures. The first one~\cite{toepfer-1971} is based on the assumption that the underlying probability distribution solving for the Vlasov equation is a Maxwellian equilibrium, which is described in the  relativistic framework by a Maxwell-J{\"u}ttner distribution~\cite{juettner-1911}. Hereafter, we will refer to it as the Local Equilibrium Closure (LEC). This choice leads to no entropy production and therefore grants the adoption of an isentropic equation of state to close for the pressure tensor field, that in this framework can be described as a single scalar 
quantity. The LEC is in principle not well suited for the description of early stage dynamics, as the restoring mechanism that drives the plasma back to an eventual initial equilibrium state (particle collisions) happens on longer time scales than the ones which are typical of the early phases of PWFA. Nevertheless, this closure would still retain a physical relevance when studying late stage dynamics (when particle collisions start to become relevant~\cite{khudiakov-2022}). Furthermore, side-by-side comparisons against other closure schemes (or PIC solvers) might indeed reveal that the LEC is also helpful for qualitative assessments on early dynamics. \\
The second closure (hereafter named warm closure - WARMC) is based on the idea proposed in~\cite{amendt-1985, newcomb-1982, siambis-1987} and later on reconsidered by~\cite{schroeder-2005, schroeder-2009, schroeder-2010}: the centralized moments equations obtained from the coarse-graining of the Vlasov equation are closed by neglecting the third order centralized moment, choice motivated by the assumption of weakly warm systems, and hence small momentum distribution variances. This leads to a closed set of equations that can be solved without having to make hypotheses on the underlying momentum probability distribution, except for the smallness of its variance. This gives also the possibility to evolve independently the various components of the pressure tensor, and hence to evaluate the expected momentum spread anisotropies: in fact, as the dynamics of the system is strongly focused in the acceleration direction, and no collisions are taken into account on these short timescales to regularize the process, momentum spread anisotropies are to be expected~\cite{newcomb-1986, amendt-1986}. \\
The main target of this paper is to develop novel numerical schemes for the simulation of warm fluid models in the context of PWFA, for both the LEC and the WARMC. The schemes rely on the lattice Boltzmann (LB) method~\cite{krueger-2017,succi-2018} to solve the fluid equations. LB is a popular numerical scheme commonly used in computational fluid dynamics as an alternate scheme to direct hydrodynamical solvers. Its formulation is rooted in the kinetic theory of gases, and this provides a strong physical basis to the method. Furthermore, the space locality of the calculations involved in the method makes this solver prone to multi CPUs and/or multi GPUs parallelization~\cite{krueger-2017}. In the past years LB has been generalized to work in many fields other than classical fluid dynamics~\cite{succi-2018}, and it is now widely accepted as a numerical solver for many physics problems that rely on a set of continuum equations. The LB formulation used in this paper
is the so-called {\it moment matching} LB, that solves systems described by advection diffusion equations~\cite{krueger-2017,succi-2018}. This is the very same formulation used in~\cite{parise-2022} to solve for the cold fluid models in PWFA. Here we extend such formulation to warm fluid models, hence we are pushing further the usage of the LB method in the context of PWFA. The LB schemes for warm fluids are coupled to a Finite Difference Time Domain (FDTD) scheme that solves for the electromagnetic fields~\cite{yee-1966,olakangil-2000}: we refer to~\cite{parise-2022} for a more in depth explanation of the  coupling. The development of the LB schemes for two different warm fluids closures will enable us to look for thermal effects that depend on the adopted closure scheme, and furthermore to assess the importance of thermal spreads anisotropies showing side-by-side comparisons between the two closures. \\
This paper is organized as follows: in Sec.~\ref{sec:lb} we provide a basic introduction to the LB method, with particular emphasis on the {\it moment matching} procedure; in Sec.~\ref{sec:relativistic} basic equations for collisionless relativistic plasmas are reviewed and the two fluid closures LEC and WARMC are discussed in Sec.~\ref{sec:lec} and Sec.~\ref{sec:warmc} respectively; numerical results will be presented in~\ref{sec:results} and an outlook and a discussion on future perspectives is given in Sec.~\ref{sec:discussion}.

\section{Lattice Boltzmann (LB) method}\label{sec:lb}

In this section, we introduce the Lattice Boltzmann (LB) method, which we use to solve the fluid equations both in the LEC and WARMC models. We first present the basics of the method, that are directly drawn from the kinetic theory of gases and in their original formulation tasked to the reproduction of classical (non relativistic) Navier-Stokes equations. We then illustrate how the whole procedure can be adapted to the simulation of generic advection equations ({\it moment matching} LB). We will see in the following sections how the relativistic warm fluid equations, in both the LEC and WARMC, can be recast into a set of advection equations, and hence can be numerically solved via {\it moment matching} LB. The interested reader might look into~\cite{krueger-2017, succi-2018} for more detailed discussions on both LB in its original formulation and its {\it moment matching} variant.

\subsection{Basic LB}\label{sec:basic-lb}
%
\begin{figure}
    \centering
    \includegraphics[width=0.7\textwidth]{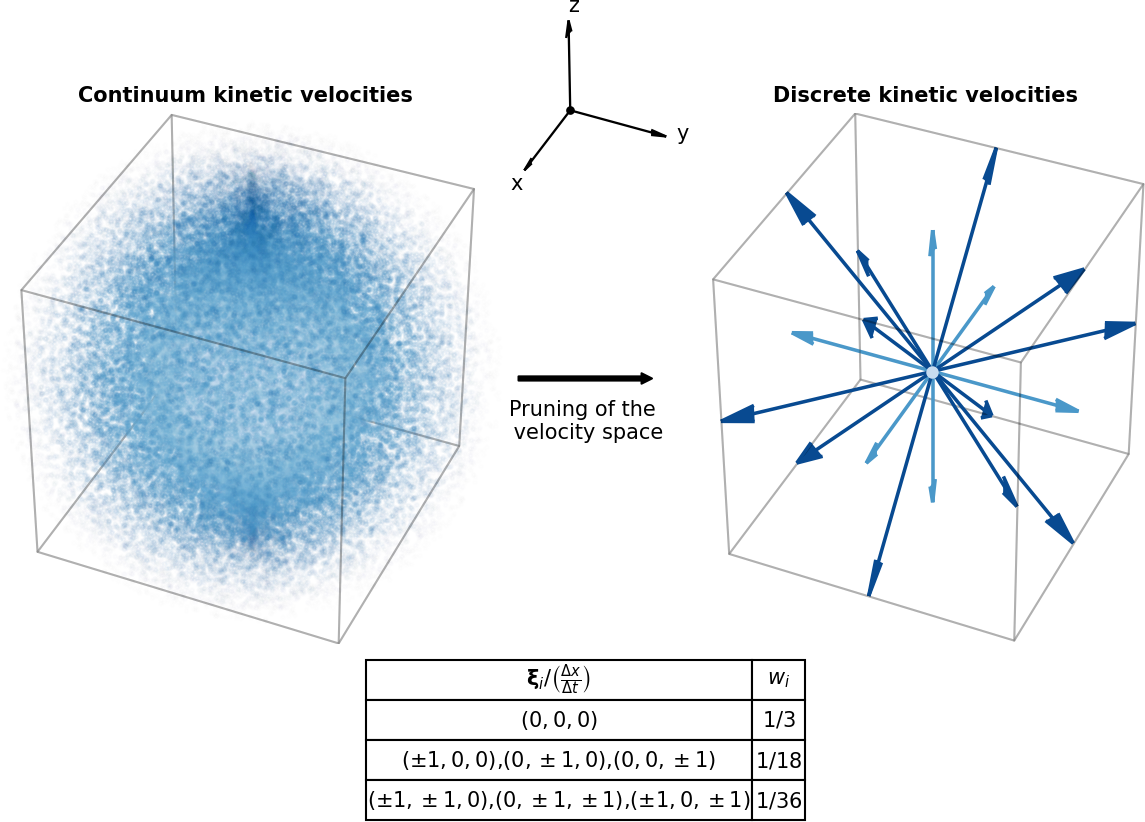}
    \caption{Lattice Boltzmann discretization. Through a pruning procedure, the continuum velocity space is discretized in a minimal set of $N_{pop}$ velocities. This discrete set (stencil) is selected in such a way as to guarantee the hydrodynamic equations for coarse grained fields (e.g. density, momentum). On the right we report one of the most popular LB stencils, with $N_{pop}=19$ velocities in 3D, which we have also used in this work. In the table we report the adimensional velocity components of the stencil, together with corresponding statistical weights. The different shades of blue group velocities with the same magnitude.}
    \label{fig:1}
\end{figure}
As the theoretical cornerstone of LB is kinetic theory, the natural starting point in its algorithmic derivation is the Boltzmann transport equation 
\begin{align} \label{eq:cont-be}
\left( \partial_t + \mybf{\xi} \cdot \nabla_{\mybf{x}} \right) f (\mybf{x}, \mybf{\xi}, t) = \Sigma (\mybf{x}, \mybf{\xi}, t) \;,
\end{align}
which describes the evolution of the phase space density $f (\mybf{x}, \mybf{\xi}, t)$ referring to the number of fluid particles with velocity $\mybf{\xi}$ at position $\mybf{x}$ at time $t$. On the RHS of the equation, $\Sigma (\mybf{x}, \mybf{\xi}, t)$ represents a production term that is usually expressed as the sum of two components: 
\begin{align}
    \Sigma (\mybf{x}, \mybf{\xi}, t) = S(\mybf{x}, \mybf{\xi}, t) + \Omega(\mybf{x}, \mybf{\xi}, t) \; ,
\end{align}
$S(\mybf{x}, \mybf{\xi}, t)$ representing the action of volume body forces $\mybf{F}$ on the particles, and $\Omega(\mybf{x}, \mybf{\xi}, t)$ the action of binary collisions:
\begin{align}
    \label{eq:source-term-cont}
    S(\mybf{x}, \mybf{\xi}, t)  &= - \mybf{F} \cdot \nabla_{\mybf{\xi}} f(\mybf{x}, \mybf{\xi}, t) \;, \\
    \Omega(\mybf{x}, \mybf{\xi}, t) &= - \frac{1}{\tau} \left( f(\mybf{x}, \mybf{\xi}, t) - f^{eq}(\mybf{x}, \mybf{\xi}, t) \right) \;,
\end{align}
where the latter has been here written by recurring to the customary Bhatnagar-Groos-Krook operator~\cite{bhatnagar-1954}, which expresses the tendency of the distribution function $f$ to relax towards an equilibrium $f^{eq}$ in a typical relaxation time $\tau$. The moments of the distribution function $f$ that solves~\cref{eq:cont-be}
\begin{align}\label{eq:cont-moments}
    M^{(n)}(\mybf{x}, t) = \int f(\mybf{x}, \mybf{\xi}, t) \mybf{\xi}^n d \mybf{\xi} \ ,
\end{align}
are then used to retrieve the hydrodynamic fields. In this basic formulation, the zeroth- and first-order moments (i.e., $n=0$, $1$) provide the mass density and the momentum density of the fluid, respectively~\cite{krueger-2017}. \\
It can then be verified through the Chapman-Enskog expansion~\cite{chapman-1970,krueger-2017,succi-2018} that the moments obtained according to~\cref{eq:cont-moments} verify the target continuum equations - again, the Navier-Stokes equations in this original formulation of the LB method - provided that $f$ is sufficiently close to $f^{eq}$. The most pivotal step in the development of the LB method is the realization that one needs only a \textit{truncated} version of the distribution functions $f$ and $f^{eq}$ to properly recover the desired moments $\mybf{M}^{(n)}(\mybf{x}, t)$ that solve the field equations~\cite{grad-1949a,grad-1949b}. For this reason, it proves expedient to expand both $f$ and $f^{eq}$ into series of orthogonal Hermite polynomials in the variable $\mybf{\xi}$, and then to truncate this expansion up to the point where one recovers the macroscopic observables of interest~\cite{shan-2006}. \\
The second most important ingredient in the development of the LB is the velocity discretization procedure. Because of the truncated Hermite polynomials expansion, one usually adopts Gauss-Hermite quadrature rules to prune the continuum velocity space and select a discrete set of $\mybf{\xi}_i$ $(i=0,\dots,N_{pop}-1)$ velocities (stencil), each of them having a statistical weight $w_i$~\cite{shan-2010,shan-2016}. In Fig.~\ref{fig:1} we report the discrete stencil adopted in this paper. This velocity discretization procedure splits~\cref{eq:cont-be} into a set of $N_{pop}$ equations, one for every discrete distribution function $f_i = f(\mybf{x}, \mybf{\xi}_i, t)$, also called \textit{population}. The careful selection of the velocity stencil is performed with the goal of \textit{exactly} preserving the coarse-graining of the moments when passing from the continuous integrals of~\cref{eq:cont-moments} to discrete summations:
\begin{align}\label{eq:disc-moments}
    M^{(n)}(\mybf{x}, t) = \int f(\mybf{x}, \mybf{\xi}, t)\mybf{\xi}^n d \mybf{\xi} 
                         = \sum_{i=0}^{N_{pop}-1} \mybf{\xi}^n_i f_i \;.
\end{align}
The performed velocity discretization, when joined with an explicit time marching discretization of step $\Delta t$, finally delivers the Lattice Boltzmann equation:
\begin{align}\label{eq:lbe}
    f_i(\mybf{x}+\mybf{\xi}_i \Delta t, t + \Delta t) = f_i(\mybf{x}, t) + \Delta t \Sigma_i(\mybf{x}, t) \; ,
\end{align}
where space is discretized on a regular lattice of characteristic length $\Delta \mybf{x} = \mybf{\xi}_i \Delta t$. The production term $\Sigma_i(\mybf{x}, \mybf{\xi}_i, t)$ is a discretization of the continuous counterpart appearing in~\cref{eq:cont-be}:
\begin{align}\label{eq:lsource}
    \Sigma_i(\mybf{x}, t) =  \Omega(\mybf{x}, \mybf{\xi}_i, t) + S_i(\mybf{x}, t) \;, 
\end{align}
with $S_i(\mybf{x}, t)$ being discretized according to one of the many forcing schemes available in the literature~\cite{krueger-2017} to properly reproduce~\cref{eq:source-term-cont}. \\ 
The algorithmic steps of the LB scheme are now clearly outlined. A set of $N_{pop}$ versions of~\cref{eq:lbe}, one for every $f_i$, is evolved on a regular spatial grid: the $f_i$ are updated at every node with the source term~\cref{eq:lsource} (\textit{Source} step), and then stream to neighbouring nodes with their corresponding velocity $\mybf{\xi}_i$ (\textit{Streaming} step). At every iteration the hydrodynamic fields are obtained through the discrete summation appearing in~\cref{eq:disc-moments}.

\subsection{Moment Matching procedure for Forced Advection Diffusion Equation}\label{sec:MMforADE}
In this section, we will make use of the framework established in Sec.~\ref{sec:basic-lb} to explain how the {\it moment matching} LB, tasked for the solution of advection diffusion equations (ADEs), works. This variant was initially developed as an alternative, more straightforward formulation of thermal LBs~\cite{shan-1997,he-1998,karlin-2013} and at the same time applied to model different physics phenomena, such as diffusive chemical reactions~\cite{dawson-1993}, combustion problems~\cite{yamamoto-2002, wang-2023}, dissolution in porous media~\cite{verhaeghe-2006,kang-2003} (more use cases can be found in~\cite{krueger-2017,succi-2018}). The target equation for a {\it moment matching} LB is a forced ADE for the scalar field $A(\mybf{x}, t)$:
\begin{align}\label{eq:ade}
\partial_t A + \nabla_{\mybf{x}} \cdot (A \mybf{u}) = \kappa \nabla_{\mybf{x}}^2 A + F \; ,
\end{align}
where the quantity $A$ is advected with velocity $\mybf{u}(\mybf{x}, t)$ and diffused via the diffusion parameter $\kappa$. The previously established framework (Sec.~\ref{sec:basic-lb}) can then be used by considering a discrete probability distribution function $f_i(\mybf{x}, t)$ whose zeroth-order discrete moment is exactly the field $A(\mybf{x}, t)$:
\begin{align}\label{eq:disc-moments-ade}
    A(\mybf{x}, t) = \sum_{i=0}^{N_{pop}-1} f_i(\mybf{x}, t) \;.
\end{align}
The velocity stencil described in Fig.~\ref{fig:1} can then be used to evolve via~\cref{eq:lbe} the distribution functions $f_i$, and the Chapman-Enskog analysis grants that the coarse graining~\cref{eq:disc-moments-ade} solves for~\cref{eq:ade}. 
The last remaining ingredient to complete the algorithmic discussion of the method is the form of the Hermite truncated expansion $f_i^{eq}$. It has been shown that a suitable form that recovers the correct equations for $A$ and minimizes the computational error is~\cite{chopard-2009}: 
\begin{align}\label{eq:ade-feq}
f^{eq}_i = w_i A \Bigg( 1 + \frac{\mybf{u} \cdot \mybf{\xi}_i}{\nu^2} + \frac{(\mybf{u} \cdot \mybf{\xi}_i)^2}{\nu^4} - \frac{\mybf{u} \cdot \mybf{u}}{2\nu^2}\Bigg)    \;,
\end{align}
where $\nu = \frac{1}{\sqrt{3}}\frac{\Delta x}{\Delta t}$ is a reference lattice velocity. This completes the description of the main algorithmic steps of the {\it moment matching} LB. The diffusion coefficient $\kappa$ is shown via Chapman-Enskog procedure to be dependent on the relaxation time $\tau$ in the following way:
\begin{align}\label{eq:diff-coeff}
    \kappa = \nu^2 \left( \tau - \frac{\Delta t}{2} \right)          \;.
\end{align}
While in this work we keep $\tau$ as small as possible (taking into account the known inferior limit $\tau > \Delta t/2$ imposed by LB bulk stability conditions~\cite{krueger-2017}) in order to recover the physics of the problem, we note from~\cref{eq:diff-coeff} that by adjusting the parameter $\tau$ in the simulations one might control the diffusive effect in the ADE, obtaining therefore a tunable regularizing effect that 
might be helpful when dealing with stiff advection equations. Finally, we report a couple of additional details for our implementation of the method. First, the source term $S_i$ is chosen to be of the form 
\begin{align}
S_i(\mybf{x}, t) = w_i F(\mybf{x}, t)   \; , 
\end{align}
in order to correctly reproduce the forcing appearing in~\cref{eq:ade}~\cite{krueger-2017,succi-2018}. Second, open boundary conditions are imposed at the extrema of the domain via the enforcing of the variables $f_i$:
\begin{align}\label{eq:open-bc}
f_i(\mybf{x}_b, t) = f_i(\mybf{x}_f, t) \; , 
\end{align}
where $\mybf{x}_b$ and $\mybf{x}_f$ are respectively the boundary position and the nearest bulk fluid node. Lastly, we remark that the whole procedure showed so far can be adapted to a 3D axisymmetric geometry: when doing so, after switching to cylindrical coordinates, we postulate that every quantity is independent from the azimuthal variable $A=A(r,z)$. Furthermore, no angular motion is requested, hence the azimuthal component of the advection velocity $\mybf{u}$ is zero. \\
Following~\cite{premnath-2005, srivastava-2013} one can develop an axisymmetric {\it moment matching} LB by considering a 2D Cartesian geometry $(r,z)$ where all the differential operators - the divergences and laplacians appearing in~\cref{eq:ade} - are carefully adapted to the cylindrical geometry~\cite{arfken-2012}. To this extent, we remark that differently from the cited sources we will consider $A$ to be either a scalar or the coordinate component of some hydrodynamic tensor field. Hence particular care will have to be taken when converting the differential operators to the cylindrical geometry. Lastly, when adopting the cylindrical geometry we have to modify the conditions applied at the $r=0$ boundary node. If the LB advected quantity $A$ is symmetric with respect to radial reflections, we keep employing~\cref{eq:open-bc}. Instead, if $A$ changes sign under radial reflections (the radial momentum component has this property) , we employ the following condition:
\begin{align}\label{eq:reflective-bc}
f_i(\mybf{x}_b, t) = f_{\overline{i}}(\mybf{x}_f, t) \; , 
\end{align}
with $\overline{i}$ being the mirrored direction to $i$ with respect to the radial axis. 

\section{Relativistic Kinetic Equations for collisionless plasma}\label{sec:relativistic}

In this section we review the basic equations that can be used to build hydrodynamic models of warm plasmas starting from the kinetic theory of gases, all expressed within the framework of special relativity~\cite{cercignani-2002, degroot-1980, rezzolla-2013}. In the next sections, we will see how the obtained set of equations can be closed via either a local equilibrium assumption or a warm closure, and then explain how to recast them into a set of advection equations. \\
In the following we will work in a flat space-time with Minkowski metric signature $\eta^{\alpha\beta}=\text{diag}(+,\mybf{-})$. When expressing formulas in a manifestly covariant form, we will adopt Einstein's summation convention, with Greek indexes running from 0 to $\rm{D}$, and Latin indexes from 1 to $\rm{D}$. When not explicitly stated, we will use $\partial_\alpha \equiv \frac{\partial}{\partial x^\alpha}$. \\
The starting point in a relativistic fluid theory of warm plasmas is the Relativistic Vlasov equation:
\begin{align}\label{eq:vlasov-eq}
\rho^{\alpha} \frac{\partial f}{\partial x^\alpha} + \frac{q}{c} F^{\alpha\beta} \rho_{\beta} \frac{\partial f}{\partial \rho^\alpha} = 0    \; ,
\end{align}
where $x^{\alpha}=(ct, \mybf{x})$ is the space-time coordinate vector, $\rho^{\alpha}=(\rho^0, \mybf{\rho})$ is the relativistic kinetic momentum vector, $F^{\alpha\beta}$ is Maxwell's electromagnetic field tensor and $f=f(x^{\alpha}, \rho^{\alpha})$ is the single-particle probability distribution function, describing the number of particles of mass $m$ and charge $q$ ($m=m_e$ and $q=-e$ in case of electrons) in the Lorentz invariant momentum space volume $\frac{d \mybf{\rho} }{\rho^0}$. \\
The fields subject of hydrodynamic theories emerge from the coarse graining of the distribution function $f$ in relativistic momentum space:
\begin{align}\label{eq:moments}
M^{\alpha_1 \cdots \alpha_n} &= c \int f \rho^{\alpha_1} \dots \rho^{\alpha_n} \frac{d \mybf{\rho}}{\rho^0} \;,
\end{align}
with the first moments being assigned the following specific names: \textit{Invariant density} $h$, \textit{Particle flow} $N^{\alpha}$ and \textit{Energy-momentum tensor} $T^{\alpha\beta}$.~\cref{eq:vlasov-eq} implies the following conservation equations for the moments~\cite{cercignani-2002,rezzolla-2013,schroeder-2010}:
\begin{align}
\label{eq:cons_eqs1}
0 &= \partial_\alpha N^{\alpha}                                           \; , \\
\label{eq:cons_eqs2}
0 &= \partial_\alpha T^{\alpha\beta} - \frac{q}{c} F^{\beta\alpha} N_{\alpha} \; , \\
\label{eq:cons_eqs3}
0 &= \partial_\alpha M^{\alpha\beta\gamma}-\frac{q}{c}(F^{\beta\alpha}T_{\alpha}^{~\gamma}+F^{\gamma\alpha}T_{\alpha}^{~\beta})                       \; . 
\end{align}
~\cref{eq:cons_eqs1,eq:cons_eqs2,eq:cons_eqs3} are the constitutive equations of a relativistic fluid, which must be coupled to Maxwell equations~\cite{jackson-1998}
\begin{align}
    \label{eq:maxwell-inhomogeneus}
    0 &= \partial_\alpha F^{\alpha\beta} - \mu_0 c q N^{\beta}  \;,  \\
    \label{eq:maxwell-homogeneus}
    0 &= \partial_\alpha F_{\beta\gamma} + \partial_\beta F_{\gamma\alpha} + \partial_\gamma F_{\alpha\beta} \;,  
\end{align}
to provide a complete description of a plasma.~\cref{eq:cons_eqs1,eq:cons_eqs2,eq:cons_eqs3} express conservation of mass, energy and momentum once a proper decomposition for $N^{\alpha}$ and $T^{\alpha\beta}$ is provided: as an example, the cold fluid equations can be obtained from~\cref{eq:cons_eqs1,eq:cons_eqs2} by assuming the following:
\begin{align}
N^{\alpha} = n_0 U^{\alpha} \;,\; T^{\alpha\beta} = n_0 m U^{\alpha}U^{\beta}   \;,
\end{align}
$U^\beta=\gamma(c, \mybf{u})$ being the fluid four velocity and $n_0$ the particle number density as measured in the fluid rest frame (density transforms under Lorentz boosts as $n=\gamma n_0$). $\gamma$ is the Lorentz factor associated to $\mybf{u}$. At variance with the cold fluid treatment, any other fluid model involving a non-zero temperature requires additional closing equations. \\
In the next sections, we will show two possible closures: the first one (LEC) postulates that the underlying distribution function describing the warm plasma (and hence solving~\cref{eq:vlasov-eq}) is a local equilibrium distribution. This grants the use of an isentropic equation of state to close for the system; the second one (WARMC) rephrases~\cref{eq:cons_eqs1,eq:cons_eqs2,eq:cons_eqs3} as centered moments equations, and then assumes the third order centered moment to be zero on the excuse of small thermal spreads, i.e. by assuming small values of the initial thermal energy ($\textit{i}$ subscripts stand for the initial values) to electron rest mass 
\begin{align}
\mu_i = \frac{k_b T_i}{m_e c^2}     \; .
\end{align}
This is the control parameter that is usually used to assess the importance of thermal effects in warm fluid models, and will be extensively used in the next sections to label our results.

\section{Local Equilibrium Closure (LEC)}\label{sec:lec}

The key assumption of the LEC hypotesis is that the distribution $f$ solving for~\cref{eq:vlasov-eq} is the relativistic version of the Maxwell-Boltzmann distribution, the Maxwell-J{\"u}ttner distribution~\cite{juettner-1911, chaconacosta-2010}. This has a number of important consequences, first and foremost the fact that under this assumption the plasma can be considered to be an ideal fluid. It can be shown~\cite{cercignani-2002} that in this case the particle flow and the energy momentum tensor assume the following forms:
\begin{align}
    N^{\alpha}      &= n_0 U^{\alpha}                                                               \; , \\
    T^{\alpha\beta} &= (P_0 + \epsilon_0) \frac{U^{\alpha}U^{\beta}}{c^2} - P_0 \eta^{\alpha\beta}  \; ,
\end{align}
where $\epsilon_0$ and $P_0$ are respectively the plasma energy density and pressure, as measured in the fluid's rest frame (all quantities written with the $0$ subscript have to be intended as they are measured in this frame). Consequently, the conservation equations~\cref{eq:cons_eqs1,eq:cons_eqs2} can be re-expressed as the relativistic counterpart of Euler's equations: 
\begin{align}\label{eq:rel-euler}
     \partial_{t} n + \nabla_{\mybf{x}} \cdot (n \mathbf{u}) &= 0 \; , \\
     \partial_t \left( \frac{h_e \mybf{p}}{m c^2}  \right) + \mybf{u} \cdot &\nabla_{\mybf{x}} \left( \frac{h_e \mybf{p}}{m c^2}  \right) = - \frac{\nabla_{\mybf{x}} P_0}{n} + q (\mybf{E} + \mybf{u} \wedge \mybf{B}) \notag \; ,
\end{align}
where $h_e = (P_0+\epsilon_0)/n_0$ is the relativistic enthalpy per particle and $\mybf{p} = m \gamma \mybf{u}$ is the relativistic fluid momentum.~\cref{eq:rel-euler} are not yet closed, but another important feature of the LEC assumption is that there is no entropy production in the fluid. Therefore to close the set of equations instead of a statement of energy conservation we consider the entropy per particle
\begin{align}
    s_0 = k_b \log \left[ \frac{1}{n_0} \left(\frac{m k_b T}{2\pi\hbar^2}\right)^{\rm{D}/2} \right] + k_b \left(1+\frac{\rm{D}}{2}\right) \;, 
\end{align}
here written in the small temperature limit~\cite{cercignani-2002}, to be constant. This statement leads to the following scaling for temperature:
\begin{align}\label{eq:temp-scaling}
    T = T_i \left(\frac{n_0}{n_i}\right)^{2/\rm{D}} \; .
\end{align}
Furthermore, this grants the use of 
the Synge equation of state~\cite{synge-1957}, that we consider here for small temperatures:
\begin{align}
    P_0        &=    n_0 k_b T = n_0 m c^2 \mu_i \left(\frac{n_0}{n_i}\right)^{2/\rm{D}}                              \; , \\               
    \epsilon_0 &= n_0 m c^2 + \frac{\rm{D}}{2}n_0 k_b T = \\
               &= n_0 m c^2 \left[ 1 + \frac{\rm{D}}{2} \mu_i \left(\frac{n_0}{n_i}\right)^{2/\rm{D}} \right]              \; . \notag
\end{align}
This provides sufficient information for the closure of~\cref{eq:rel-euler}, and paves the way to a successful numerical treatment of this set via {\it moment matching} LB. 

\subsection{Local Equilibrium Closure Lattice Boltzmann (LEC-LB)}\label{sec:lec-lb}

We can now express~\cref{eq:rel-euler} as a series of forced advection equations:
\begin{align}\label{eq:lb-ade-lec}
     \partial_t \mybf{A} + \nabla_{\mybf{x}} \cdot (\mybf{u} \mybf{A}) \equiv  \mathcal{D}_{\mybf{u}} (\mybf{A}) = \mybf{F}        \;,  
\end{align}
where
\begin{align} \label{eq:lb-ade-lec-comps-A}
    \mybf{A} &= 
    \begin{pmatrix}
        n               \\
        \left[ 1 + \left( 1 +\frac{\rm{D}}{2} \right) \mu_i \left(\frac{n}{\gamma n_i}\right)^{2/\rm{D}}  \right] n \mybf{p}
    \end{pmatrix}   \; , 
\end{align}
\begin{align} \label{eq:lb-ade-lec-comps-F}
    \mybf{F} &= 
    \begin{pmatrix}
        0               \\
        - n_i m c^2 \mu_i \nabla_{\mybf{x}} \left(\frac{n}{\gamma n_i}\right)^{1+2/\rm{D}} + q n (\mybf{E} + \mybf{u} \wedge \mybf{B})
    \end{pmatrix}   \; .    
\end{align}
This set can then be numerically solved via {\it moment matching} LB, using the techniques explained in Sec.~\ref{sec:MMforADE}. As already stated, we can adapt the method to work in a 3D axisymmetric environment, and particular care has to be taken when translating the equations to said geometry. As the final expressions are rather bulky, we report them in full in Appendix~\ref{app:full-eqs}. \\ 
The two-way coupling of the fluid with the electromagnetic fields needs no particular explanation: the plasma hydrodynamic quantities (particle density $n$ and the transport velocity $\mybf{u}$) are obtained from the LB evolved quantities (the various components of $\mybf{A}$) and then fed to the FDTD Maxwell solver together with the driving bunch terms. The evolved electromagnetic fields are then plugged into~\cref{eq:lb-ade-lec-comps-F} as source terms, and the next LB iteration can be performed. \\
The only detail worth of discussion is the determination of the transport velocity $\mybf{u}$ from the LB-advected quantities $\mybf{A}$. In fact, due to the appearance of the $\gamma$ Lorentz factor in~\cref{eq:lb-ade-lec-comps-A}, one cannot easily obtain the relationship between $\mybf{u}$ and the second component of the vector $\mybf{A}$ (the one containing momentum $\mybf{p})$, and therefore there is the need for a specific iterative algorithm to determine it. 
Initially set $\mybf{p}^{(0)}$ as the zero-th order moment of the distribution function coming out of the LB iteration, divided by $n$:
Then:
\begin{enumerate}
    \item Starting from $k=0$, compute $\mybf{u}^{(k)}$ as 
    \begin{align}
        \mybf{u}^{(k)} = \frac{\mybf{p}^{(k)}}{m \sqrt{1+|\mybf{p}^{(k)}/(mc)|^2}}    \; .
    \end{align}
    \item Compute $\gamma^{(k)}=\gamma(\mybf{u}^{(k)})$ as 
    \begin{align}
        \gamma^{(k)} = \gamma(\mybf{u}^{(k)}) = \frac{1}{\sqrt{1-(\mybf{u}^{(k)}/c)^2}} \; .
    \end{align}
    \item Set $\mybf{p}^{(k+1)}$ as
    \begin{align}
        \mybf{p}^{(k+1)} = \frac{\mybf{p}^{(0)}}{1 + \left( 1 +\frac{\rm{D}}{2} \right) \mu_i \left(\frac{n}{\gamma n_i}\right)^{2/\rm{D}}} \; .
    \end{align}
    \item Repeat until convergence is reached.
\end{enumerate}
\section{Warm Plasma Closure (WARMC)} \label{sec:warmc}

The derivation given in this section closely follows~\cite{schroeder-2010}. To explain how to derive the Warm Closure from the conservation equations~\cref{eq:cons_eqs1,eq:cons_eqs2,eq:cons_eqs3} one has to first define the \textit{thermal momentum} $w^{\mu}$ as 
\begin{align}\label{eq:thermal-momentum}
    w^{\mu} = \left( \frac{n_0}{h} \right) U^{\mu} \quad \Rightarrow \quad N^{\mu} = n_0 U^{\mu} = h w^{\mu}    \; .
\end{align} 
The second and third order $\textit{centered}$ moments of the distribution function $f$, respectively $\theta^{\mu\nu}$ and $Q^{\mu\nu\lambda}$, are then defined as:
\begin{align}
     \theta^{\mu\nu} &= c \int (\rho^\mu - w^\mu) (\rho^\nu - w^\nu) f \frac{d \mybf{\rho}}{\rho^0}                                         \;, \\
          Q^{\mu\nu\lambda} &= c \int (\rho^\mu - w^\mu) (\rho^\nu - w^\nu) (\rho^\lambda - w^\lambda) f \frac{d \mybf{\rho}}{\rho^0}       \;. 
\end{align} 
Therefore one obtains:
\begin{align}\label{eq:derive-central-mom}
     T^{\mu\nu} &= \theta^{\mu\nu} + h w^\mu w^\nu                                                  \;, \\
     M^{\mu\nu\lambda} &= Q^{\mu\nu\lambda} + w^\mu     \theta^{\nu\lambda} 
                                                 + w^\nu     \theta^{\mu\lambda}
                                                 + w^\lambda \theta^{\mu\nu} 
                                                 + h w^\mu w^\nu w^\lambda                          \;, 
\end{align}
and due to the mass-shell condition $\rho^\mu \rho_\mu = m^2 c^2$ it follows that:
\begin{align}
    \label{list:traces1}
    T^\mu_{~\mu}        &= h m^2 c^2 \hspace{0.33cm} \Rightarrow \theta^\mu_{~\mu}=h(m^2c^2-w^\mu w_\mu)  \;, \\
    \label{list:traces2}     
    M^{\mu\nu}_{~~~\nu} &= N^\mu m^2c^2              \Rightarrow Q^{\mu\nu}_{~~~\nu}=2w_\nu\theta^{\mu\nu}   .       
\end{align}
The conservation equations~\cref{eq:cons_eqs1,eq:cons_eqs2,eq:cons_eqs3} become therefore
\begin{align}
      0 &= \partial_\alpha (h w^\alpha)                                                                     \;, \\
      0 &= \partial_\alpha \theta^{\alpha\beta} 
         + \partial_\alpha ( h w^\alpha w^\beta ) 
         - \frac{h q}{c}F^{\beta\alpha}w_{\alpha}                                                           \;, \\
      0 &= \partial_\alpha Q^{\alpha\beta\gamma}
         + \partial_\alpha \left( h w^\alpha \frac{\theta^{\beta\gamma}}{h} \right) 
         + \theta^{\gamma\alpha} \partial_\alpha w^\beta
         + \theta^{\alpha \beta} \partial_\alpha w^\gamma \notag \\
         &- \frac{q}{c}(F^{\beta\alpha}\theta_{\alpha}^{~\gamma}+F^{\gamma\alpha}\theta_{\alpha}^{~\beta})   \;.
\end{align}
The warm closure consists in taking $Q^{\alpha\beta\gamma} = 0$, based on the assumption of a small momentum spread:
\begin{align}
      \partial_\alpha (n_0 U^\alpha) &= 0                                                                     \label{eq:warmc-fluid-eq1}\;, \\
      \partial_\alpha \left( n_0 U^\alpha \frac{n_0 U^{\beta}}{h} \right) &= 
         - \partial_\alpha \theta^{\alpha\beta} 
         + \frac{n_0 q}{c}F^{\beta\alpha}U_{\alpha}                                                           \label{eq:warmc-fluid-eq2}\;, \\
      \partial_\alpha \left( n_0 U^\alpha \frac{\theta^{\beta\gamma}}{h} \right) &=  
         - \theta^{\gamma\alpha} \partial_\alpha \left( \frac{n_0 U^{\beta}}{h} \right) 
         - \theta^{\alpha \beta} \partial_\alpha \left( \frac{n_0 U^{\gamma}}{h} \right) + \notag \\
         &+ \frac{q}{c}(F^{\beta\alpha}\theta_{\alpha}^{~\gamma}+F^{\gamma\alpha}\theta_{\alpha}^{~\beta})    \label{eq:warmc-fluid-eq3}\;,
\end{align}
and this also provides an additional condition, obtained from~\cref{list:traces2}:
\begin{align}
    \label{eq:traces3}
    \theta^\mu_{~\mu} &= h c^2 \left[ m^2 - \left(\frac{n_0}{h}\right)^2 \right] \; , \\
    \label{eq:traces4}
    U_\nu \theta^{\mu\nu} &= 0                 \; .
\end{align}
Note that the previous expressions have been written using~\cref{eq:thermal-momentum} to express everything in terms of the more familiar quantities $n_0$ and $U^{\alpha}$.~\cref{eq:warmc-fluid-eq1,eq:warmc-fluid-eq2,eq:warmc-fluid-eq3,eq:traces3,eq:traces4} constitute a closed set of fluid equations, which may be solved for the plasma evolution through {\it moment matching} LB.

\subsection{Warm Plasma Closure Lattice Boltzmann (WARMC-LB)} \label{sec:warmc-lb}
~\cref{eq:warmc-fluid-eq1,eq:warmc-fluid-eq2,eq:warmc-fluid-eq3,eq:traces3,eq:traces4} can be 
rephrased as advection equations by realizing that, for any given quantity $G=\left\{1, n_0 U^{\beta}/h, \theta^{\beta\gamma}/h \right\}$, one can write:
\begin{align}
    \partial_\alpha(n_0 U^\alpha G) = \partial_t(n G) + \nabla_{\mybf{x}} \cdot (nG\mybf{u}) = \mathcal{D}_{\mybf{u}}(nG)  \; .
\end{align}
Therefore one has:
\begin{align}
     \partial_t \mybf{A} + \nabla_{\mybf{x}} \cdot (\mybf{u} \mybf{A}) \equiv  \mathcal{D}_{\mybf{u}} (\mybf{A}) = \mybf{F}        \;,  
\end{align}
where
\begin{align} \label{eq:lb-ade-warmc-comps-A}
    \mybf{A} &= 
    \begin{pmatrix}
        n                                   \\
        n \frac{n_0 U^{\beta}}{h}           \\
        n \frac{\theta^{\beta\gamma}}{h}
    \end{pmatrix}   \; ,
\end{align}
\begin{align} \label{eq:lb-ade-warmc-comps-F}
    \mybf{F} &= 
    \begin{pmatrix}
        0                                                                                   \\
        - \partial_\alpha \theta^{\alpha\beta} + \frac{n_0 q}{c}F^{\beta\alpha}U_{\alpha}   \\
        {\scriptstyle - \theta^{\gamma\alpha} \partial_\alpha \left( \frac{n_0 U^{\beta}}{h} \right) 
        - \theta^{\alpha \beta} \partial_\alpha \left( \frac{n_0 U^{\gamma}}{h} \right) +
        + \frac{q}{c}(F^{\beta\alpha}\theta_{\alpha}^{~\gamma}+F^{\gamma\alpha}\theta_{\alpha}^{~\beta})}
    \end{pmatrix}  \; .    
\end{align}
The system of equations can now be simulated via {\it moment matching} LB, interpreting all terms on the RHS of equations as external forcings. The Fluid-Maxwell coupling discussed in Sec.~\ref{sec:lec-lb} applies also in this case, with the exception of the iterative scheme for the transport velocity $\mybf{u}$, since in this framework $\mybf{u}$ can be naturally identified from the LB advected quantities. \\
Also in this case we employ the axisymmetric description: the passages that are needed to adapt \cref{eq:lb-ade-warmc-comps-A,eq:lb-ade-warmc-comps-F,eq:traces3,eq:traces4} to this framework are rather lengthy but simple and we report the final full expressions in Appendix~\ref{app:full-eqs}. \\ 
As a final remark, it can be worth mentioning how the tensor $\theta^{\alpha\beta}$ can be decomposed in its transversal and longitudinal pressure components (w.r.t.~the direction of the driving bunch, here chosen to be the $z-$axis), respectively $P_{\perp}$ and $P_{\parallel} = P_{\perp} + \Delta P$, as to define a term of comparison against the isotropic pressure case provided by the LEC. We postulate~\cite{rezzolla-2013,letelier-1980} that in the fluid rest frame the anisotropic energy momentum tensor and the thermal momentum would read as 
\begin{align}                                    
    T^{\mu\nu}_0 &= \text{diag}(\epsilon_0, P_\perp, P_\perp, P_\parallel)         \;, \\
    w^{\mu}_0 &= \frac{n_0}{h}(c,0,0,0)                                          \;.
\end{align}
It follows then from~\cref{eq:derive-central-mom} that 
\begin{align}
    \theta^{\mu\nu}_0 = \text{diag}\left(\epsilon_0-\frac{n_0^2 c^2}{h}, P_{\perp}, P_{\perp}, P_{\perp}\right) + \Delta P \delta^{\mu}_{~z}\delta^{\nu}_{~z}\;. 
\end{align}
It is sufficient to apply a Lorentz boost $\Lambda^{\mu}_{~\nu}$ to derive this tensor in a generic
reference frame:
\begin{align}\label{eq:double-boost}
    \theta^{\mu\nu} &= \Lambda^{\mu}_{~\alpha} \Lambda^{\nu}_{~\beta} \theta^{\alpha\beta}_{0} = \\
                    &= \left( P_\perp + \epsilon_0 - \frac{n_0^2c^2}{h} \right) \frac{U^\mu U^\nu}{c^2} - P_\perp \eta^{\mu\nu} + \Delta P \Lambda^{\mu}_{~z} \Lambda^{\nu}_{~z} \;. \notag
\end{align}
Other than showing how to obtain the rest frame values for both pressure terms, this form of the 
tensor is actually useful to determine that some of its components are null in the axisymmetric framework. This effectively reduces the number of advection equations that have to be solved via {\it moment matching} LB.

\section{Numerical Results}\label{sec:results}
%
\begin{figure}
    \centering
    \includegraphics[width=0.8\textwidth]{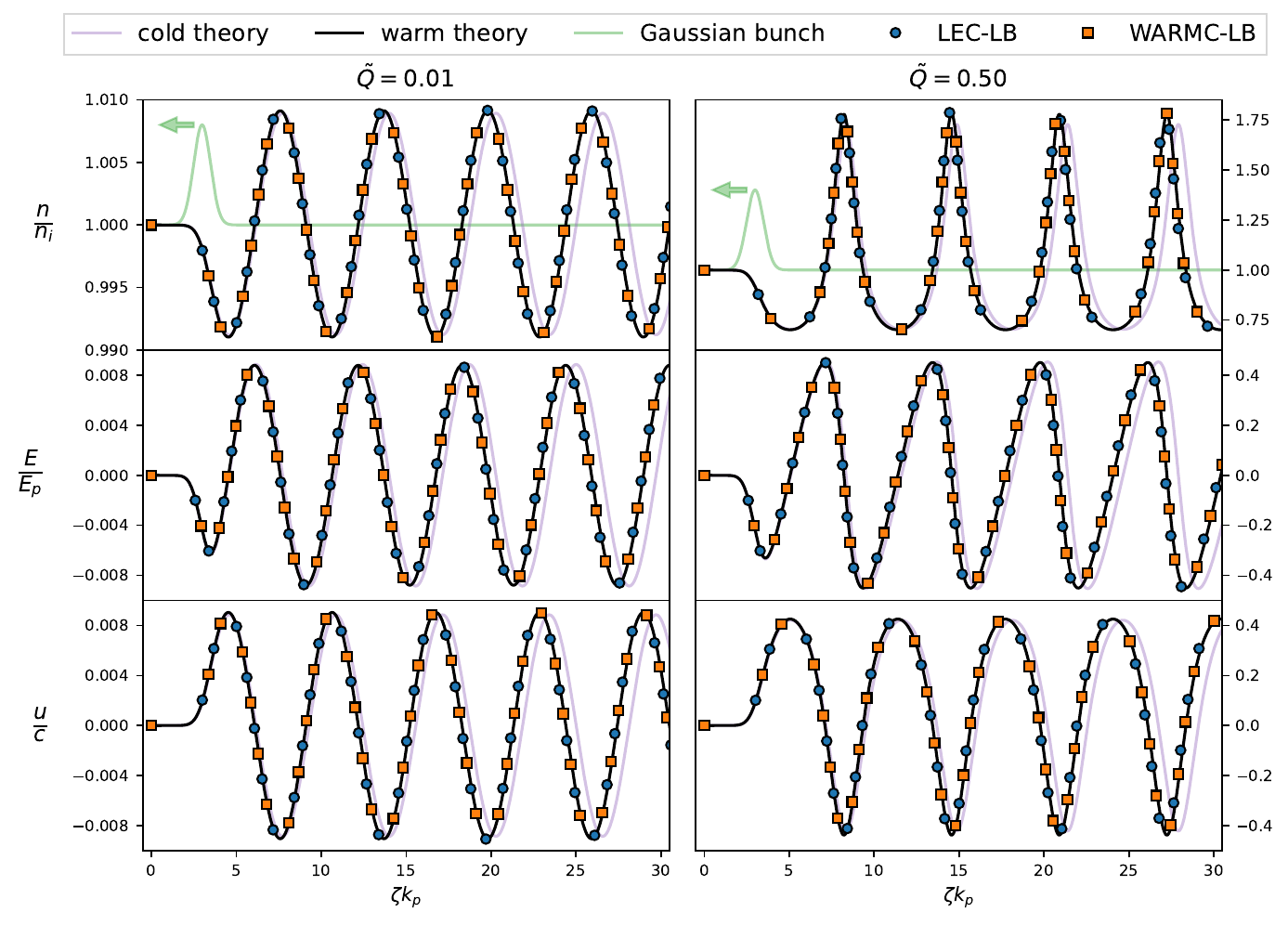}
    \caption{Comparison between the numerical results obtained from the {\it moment matching} LB with the LEC (LEC-LB), the {\it moment matching} LB with the WARMC (WARMC-LB) and their respective warm ($\mu_i = 0.04 \rightarrow k_b T_i = 20$ $\rm{KeV}$) 1D theory~\cite{rosenzweig-1988,schroeder-2005} (which coincide for this choice of the parameters and hence is represented here by a single solid black line). We also show for reference the analytic solution that can be derived from the cold fluid model (light purple). 
    From top to bottom, we show results for the electron plasma density $n$, the electric field $E$ (normalized w.r.t.~$E_p=\frac{m c \omega_p}{e}$), and the plasma velocity $u$. All the curves are plotted with the co-moving variable $\zeta=z-ct$ on the $x-$axis. We show two significant regimes: $\tilde{Q}=0.01$ on the left column and $\tilde{Q}=0.5$ on the right column. The Gaussian driving bunch (green line) appearing in the top panels is vertically shifted by a unit factor for visualization purposes.}
    \label{fig:2}
\end{figure}
In this section we present the current capabilities of the method by reproducing known analytical results and we also show side-by-side comparisons between the two closures, highlighting the emergence of momentum spreads anisotropies. We divide the showcase of our results in three different parts: in the first we start by considering temperature effects in a completely 1D scenario, where our equations of motions are made mono-dimensional by imposing translational symmetry along the radial directions (all derivatives w.r.t. transversal coordinates are therefore zero), imposing $\rm{D}=1$ in~\cref{eq:lb-ade-lec-comps-A,eq:lb-ade-lec-comps-F} (this is equivalent to considering a 1D kinetic momentum space) and considering $(1+1)-$dimensional tensors in the WARMC case. In this 1D1V set-up, we compare our numerical result with known analytical solutions~\cite{rosenzweig-1988,katsouleas-1988,schroeder-2005,schroeder-2009}. Then, we move to the discussion of dispersion relations in a 1D3V setup~\cite{clemmow-1956,schroeder-2010}: there is still translational invariance along the transversal directions, but a 3D kinetic momentum space is considered ($\rm{D}=3$ in~\cref{eq:lb-ade-lec-comps-A,eq:lb-ade-lec-comps-F} and $\theta^{\mu\nu}$ tensors are (3+1) dimensional).
Lastly, we consider full spatially resolved simulations in a 3D axisymmetric environment (3D3V). \\
In this work, plasma ions are considered as an immobile background with constant plasma density $n_i=10^{16}$ $\rm{cm}^{-3}$, as they are several order of magnitude more massive than plasma electrons. Furthermore, the driving electron pulse is modelled by a rigid bi-Gaussian density $n_b$, with rms-sizes $\sigma_{z}=\sigma_{r}=25$ $\rm{\mu m}$ and moving at the speed of light $c$ (moving from right to left):
\begin{align}\label{eq:bunch-density}
n_b(r,z) = \alpha \exp{\left(-\frac{(z-ct-z_0)^2}{\sigma_z^2}\right)} \exp{\left(-\frac{r^2}{\sigma_r^2}\right)} \; .
\end{align}
The driving bunch is initially centered in $z_0$ and with a peak amplitude $\alpha$ such that the \textit{Normalized Charge Parameter} $\tilde{Q}$
\begin{align}\label{eq:Qtilde}
\tilde{Q} = (2\pi)^{\rm{D}/2} \sigma_z \sigma_r^{\rm{D}-1} \alpha \left( \frac{k_p^{\rm{D}}}{n_i} \right)    \; ,
\end{align}
representing the strength of the perturbation, is kept at a fixed desired value. Here $k_p$ is the plasma wave number, $k_p=\omega_p/c$, with $\omega_p=\sqrt{e^2 n_i / (m \epsilon_0)}$ the cold plasma frequency ($m$ being the electron mass and $\epsilon_0$ the vacuum permittivity). Unless explicitly stated, all the results presented from now on will consider a computational domain of size $L_r = 6 / k_p$ and $L_z =30 / k_p$, with cell resolutions $\Delta z = \Delta r = 0.01/ k_p$ and computational time step $\Delta t = 0.001 / \omega_p$, for a simulated physical time of $30 / {\omega_p}$. The value of $\tau$ is tuned in every setup in order to gauge between the LB numerical stability conditions~\cite{krueger-2017} and the smallness of the parameter $\kappa$ introduced in~\cref{eq:diff-coeff}, as to properly recover the physics of the problem: we have chosen a value of $\tau=0.52\Delta t$ in the 1D setups, and $\tau=0.53\Delta t$ in the 3D setups, and both these two values perfectly reproduce the available analytic solutions.\\
As already mentioned, just like other common PWFA solvers~\cite{marocchino-2016} the electromagnetic fields are solved via an FDTD scheme~\cite{yee-1966,olakangil-2000} through numerical integration of the curl Maxwell equations (Faraday's and Ampere's laws in~\cref{eq:maxwell-inhomogeneus,eq:maxwell-homogeneus}). It can be shown in fact~\cite{massimo-2015} that when these equations are considered together with the continuity equation, the divergence Maxwell counterparts (Gauss laws in~\cref{eq:maxwell-inhomogeneus,eq:maxwell-homogeneus}) are automatically satisfied, provided that the fields are correctly initialized. For this reason, we properly initialize the electromagnetic fields by solving analytically the Maxwell system with just a rigid Gaussian density (the driving bunch) in its rest-frame, and then boost the quantities to the lab-frame~\cite{londrillo-2014,massimo-2016}. 

\subsection{1D1V results}

We adapt our scheme to work in a 1D environment by employing a value of $\sigma_r$ which is way bigger than the computational domain along the radial direction $L_r$, so that the driving bunch of~\cref{eq:bunch-density} reduces effectively to a single Gaussian profile along the $z$ coordinate, and the system assumes a translational invariance along the $r$ coordinate.~\cref{eq:Qtilde} is therefore employed by imposing $\rm{D}=1$. Furthermore, the 1V condition is reached for the LEC by setting $\rm{D}=1$ in~\cref{eq:lb-ade-lec-comps-A,eq:lb-ade-lec-comps-F}, while in the WARMC case is sufficient to initialize to zero the transversal components of the $\theta^{\mu\nu}$ tensor. \\
We present in Fig.~\ref{fig:2} the results for a numerical benchmark of the method obtained by comparing both the two fluid solvers against the analytic solutions that can be obtained by considering~\cref{eq:rel-euler} and~\cref{eq:warmc-fluid-eq1,eq:warmc-fluid-eq2,eq:warmc-fluid-eq3,eq:traces3,eq:traces4} in a 1D1V environment. The corresponding equations are well studied in the literature: for the LEC see Sec.~VI in~\cite{rosenzweig-1988} or equivalently~\cite{katsouleas-1988}; for the WARMC see~\cite{schroeder-2005}, and replace the laser driven case with a particle driven case. The theoretical solutions are obtained via finite differences integration of Eq.~(6.4) in~\cite{rosenzweig-1988} (LEC) and Eq.~(8) in~\cite{schroeder-2005} (WARMC). Although coming from different set of equations, these solutions are equivalent in the limit of small temperature. Being this the case, they are represented by a single solid black curve in Fig.~\ref{fig:2}. We observe that both the two methods are perfectly able to reproduce the theoretical results. In this simplified 1D scenario it is also possible to start appreciating some effects of temperature on the dynamics (the choice of $\mu_i$ is done in order to evidence thermal effects in the first periods of the wave): by comparing against the analytic solutions of the cold fluid model (light purple in Fig.~\ref{fig:2}) it is possible to see that temperature leads to a decrease in the thermal wavelength of the plasma wave. As it will be seen in the next section, this effect is in stark contrast to what can be observed in a 3V environment, where the wavelength increases or decreases depending on the selected fluid closure. 

\subsection{1D3V results}

\begin{figure}
    \centering
    \includegraphics[width=0.5\textwidth]{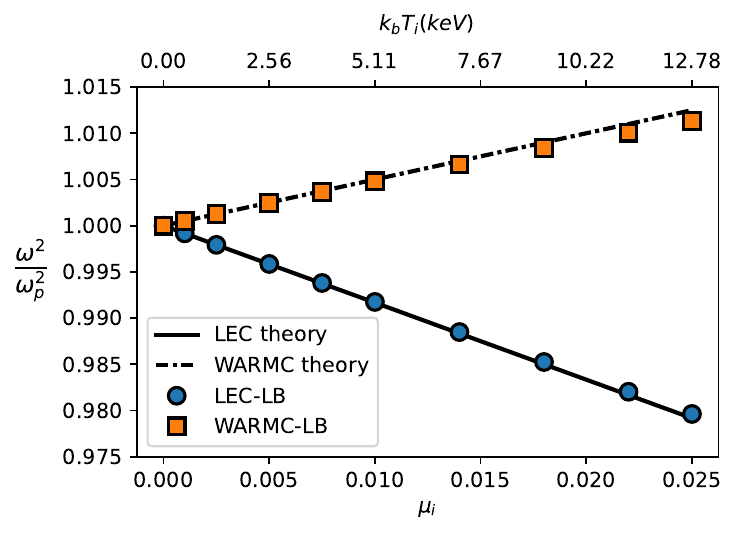}
    \caption{Dispersion relation dependency on temperature. Comparison of the numerics obtained from the {\it moment matching} LB simulations versus the theory predictions~\cref{eq:disp-rel1,eq:disp-rel2}. In both the two closures, our numerical codes well reproduce the theoretical results. We remark that theory predictions are operated in the linear regime (small $\tilde{Q}$ perturbation) and in the assumption of small initial temperatures (small $\mu_i$ values).}
    \label{fig:3}
\end{figure}

An increase in complexity with respect to the 1D1V case is represented by the 1D3V case, that is targeted specifically at the recovery of dispersion relations for the plasma wave, while still retaining a 1D simulation framework. In this context it is actually possible to extract for both closures a dispersion relation linking the frequency $\omega$ of the plasma wave to its wavenumber $\mybf{k}$. In the literature, such result has been obtained with various levels of approximations~\cite{bohm-1949,clemmow-1956}. In our case, a theoretical study can be conducted by considering~\cref{eq:rel-euler} and~\cref{eq:warmc-fluid-eq1,eq:warmc-fluid-eq2,eq:warmc-fluid-eq3,eq:traces3,eq:traces4}, linearly perturbing them and then obtaining the dispersion relation via Fourier analysis~\cite{schroeder-2010,hutchinson-2003,lee-2023}. 
The following formulas are obtained at order $O(\mu_i)$:
\begin{align}
    \left(\frac{\omega}{\omega_p}\right)^2 &= 
    1 - \frac{5}{2} \mu_i + \frac{5}{3} \mu_i \left(\frac{\mybf{k}}{k_p} \right)^2 
    \quad \text{LEC}        \; , \\
    \left(\frac{\omega}{\omega_p}\right)^2 &= 
    1 - \frac{5}{2} \mu_i + 3 \mu_i \left(\frac{\mybf{k}}{k_p} \right)^2 
    \quad \text{WARMC}      \; . 
\end{align}
Note that the WARMC result in the previous equation coincides with findings in~\cite{clemmow-1956} and represents the relativistic Bohm-Gross relationship for warm Langmuir waves, where the $5/2$ correction term stands for thermal inertia  i.e. hydrodynamic transport correction due to temperature. \\
The 1D3V framework is such that only longitudinal waves are present in the system, and therefore one has phase velocity equal to the bunch velocity $v_{ph}=\frac{\omega}{k}=c$. Therefore, in this context one obtains at order $O(\mu_i)$
\begin{align}
    \label{eq:disp-rel1}
    \left(\frac{\omega}{\omega_p}\right)^2 &= 1 - \frac{5}{6} \mu_i 
    \quad \text{LEC}        \; , \\
    \label{eq:disp-rel2}
    \left(\frac{\omega}{\omega_p}\right)^2 &= 1 + \frac{1}{2} \mu_i  
    \quad \text{WARMC}      \; . 
\end{align}
At variance with what was observed in the previous section,~\cref{eq:disp-rel1} foresees a closure dependent behavior of the wave: with respect to the cold case (where $\omega=\omega_p$), LEC plasmas exhibit wavelengths decreasing with temperature, whereas WARMC plasmas wavelengths increase with temperature. This can be observed in Fig.~\ref{fig:3}, where we report a comparison between the theoretical predictions and the numerical results. In both the two closure schemes we observe a very good agreement.
%

%

\subsection{3D3V results}
\begin{figure}
    \centering
    \includegraphics[width=\textwidth]{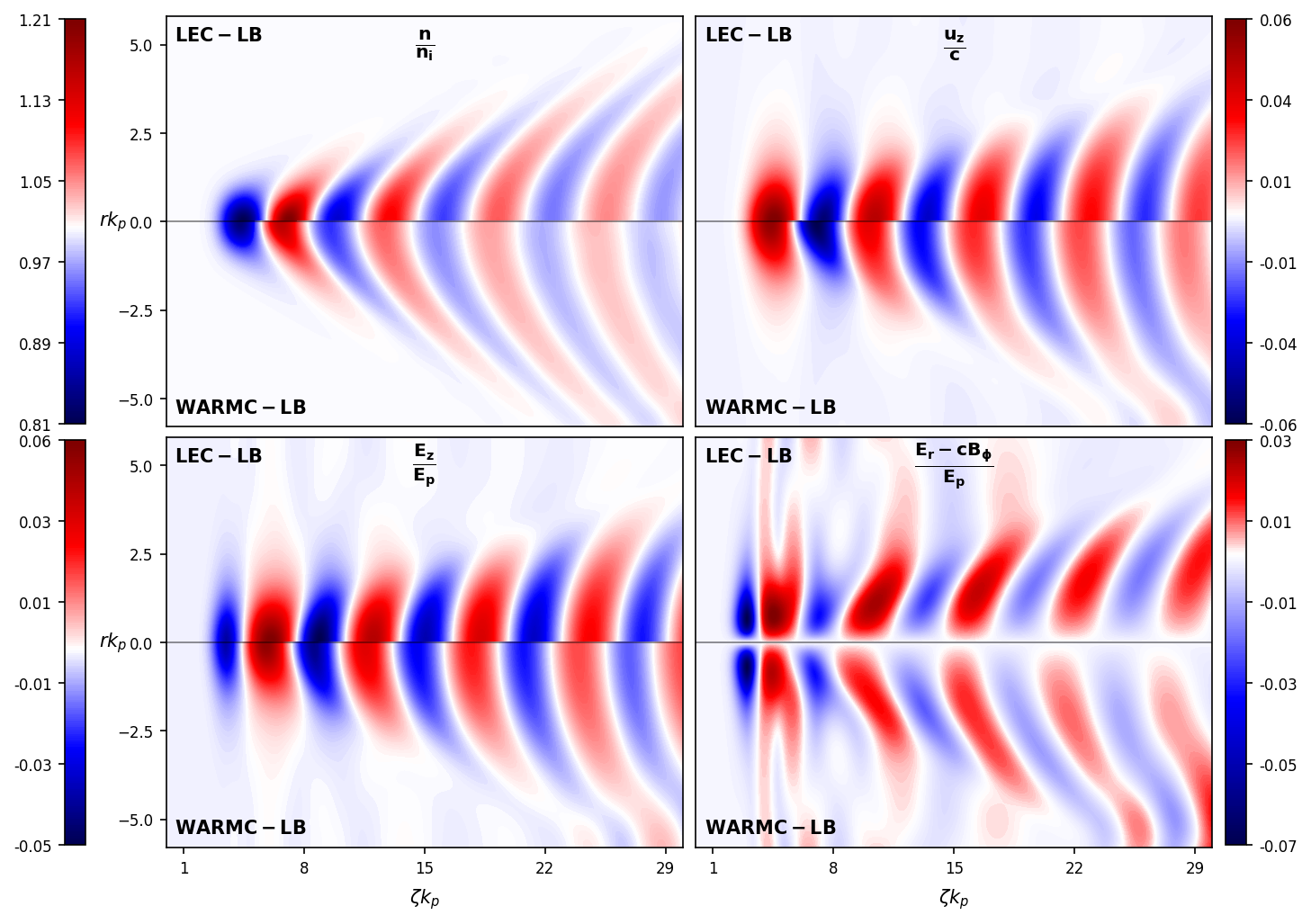}
        \caption{Qualitative comparison of 3D3V LEC-LB simulation against the WARMC-LB counterpart. We show four significant quantities, all properly adimensionalized w.r.t.~previously defined values: the particle density $n$, the longitudinal fluid velocity $u_z$ and both the two components of the wakefields, $E_z$ and $E_r-c B_\phi$. The chosen initial temperature value is $\mu_i = 0.04 \rightarrow k_b T_i = 20$ $\rm{keV}$ (again, selected for enhancing thermal effects in the first wave periods) while $\tilde{Q}=0.5$. }
    \label{fig:4}
\end{figure}

\begin{figure}
    \centering
    \includegraphics[width=0.5\textwidth]{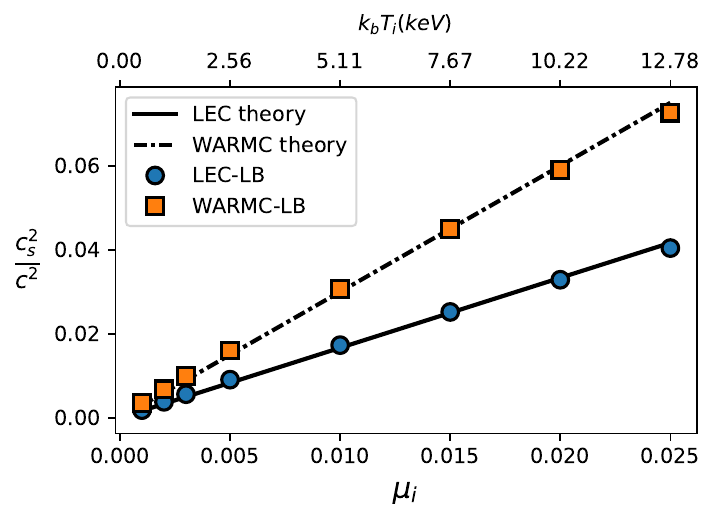}
    \caption{Acoustic velocity $c_s$ dependency on the initial plasma temperature $\mu_i$. Comparison of the numerics obtained from the {\it moment matching} LB simulations versus the warm linear theory predictions~\cref{eq:sound-speed1,eq:sound-speed2}. In both the two closures, our numerical codes well reproduce the theoretical results. We remark that theory predictions are operated in the linear regime (small $\tilde{Q}$ perturbation) and in the assumption of small initial temperatures (small $\mu_i$ values).}
    \label{fig:6}
\end{figure}

\begin{figure}[b]
    \centering
    \includegraphics[width=\textwidth]{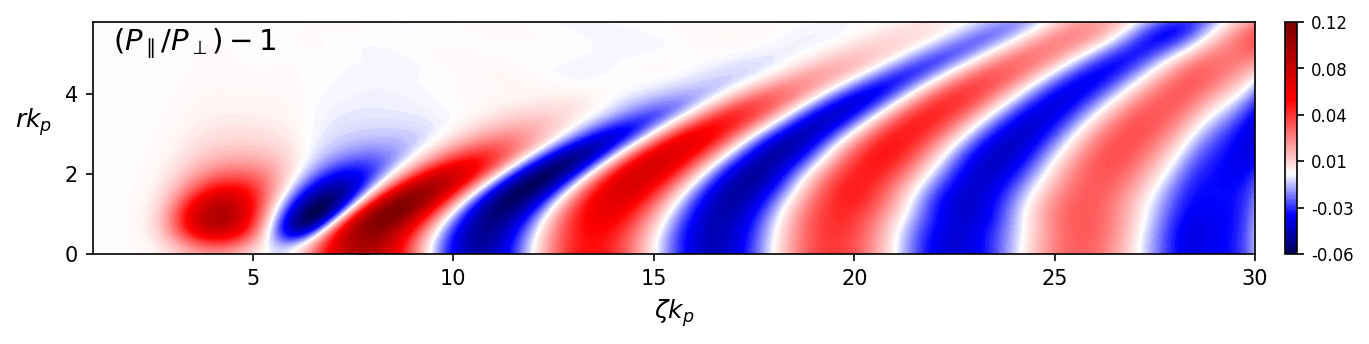}
    \caption{Study of the thermal spread anisotropies in the WARMC model. The pressure ratio $P_{\parallel}/P_{\perp}$, whose value is obtained in the numerics via inversion of~\cref{eq:double-boost}, is compared against its isotropic value $1$. Differences of order $10\%$ are observed in proximity of the bunch. The chosen initial temperature value is $\mu_i = 0.04 \rightarrow k_b T_i = 20$ $\rm{KeV}$ (again, selected for enhancing thermal effects in the first wave periods) while $\tilde{Q}=0.5$.}
    \label{fig:7}
\end{figure}

We now move our analysis to fully spatially resolved simulations. In Fig.~\ref{fig:4} we provide an initial comparison of the two fluid closures in a 3D3V axisymmetric environment. The temperature is set to an initial value of $k_b T_i = 20$ $\rm{keV}$, and the chosen regime of the driving pulse is quasi-linear, $\tilde{Q}=0.5$. In the figure, we show color plots for four significant quantities: the plasma number density $n$, the longitudinal fluid velocity $u_z$, the longitudinal accelerating field $E_z$ and the transversal wakefield $E_r - c B_\phi$. It is possible to appreciate that at variance with the usual dynamics of the cold case~\cite{parise-2022}, the typical wave-like pattern of peaks and valleys is perturbed by the emergence of an acoustic behavior, that promotes electron motion out of the radial axis. The presence of pressure waves can be appreciated by the appearance of Mach cone structures~\cite{landau-1987}, that can be easily spotted by looking at anyone of the panels of Fig.~\ref{fig:4}: as the driving bunch travels at the speed of light $c$, it perturbs the plasma medium. The perturbation propagates at a specific velocity $c_s$ (that depends on the initial temperature $\mu_i$) smaller than $c$, and the wave-fronts of the perturbation, that in a sub-sonic flow would radially propagate all over the space, are instead constrained in a conical structure around the perturbation: a super-sonic Mach cone is observed. \\
\begin{figure}
    \centering
    \includegraphics[width=\textwidth]{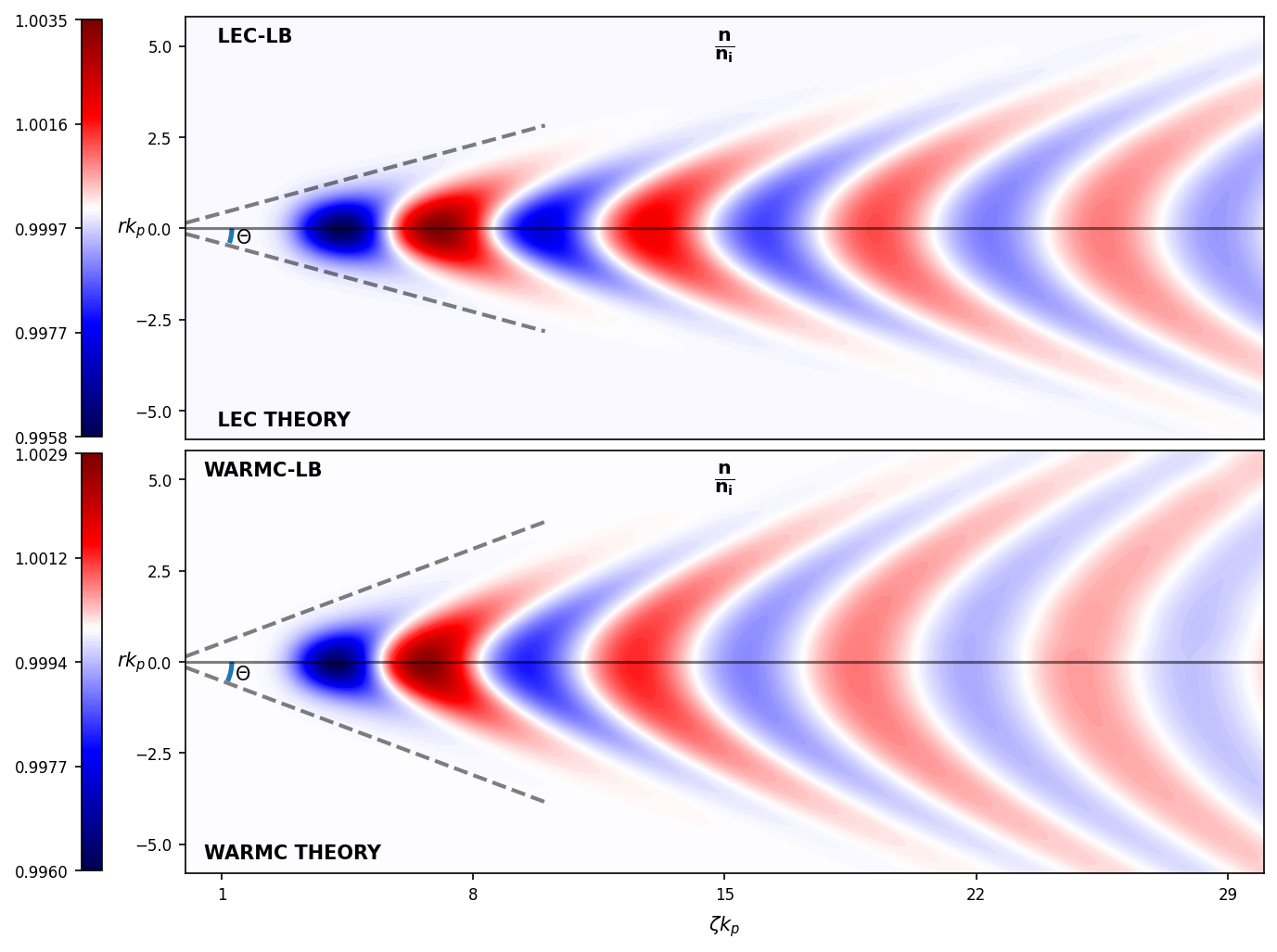}
    \caption{Comparison of simulations (top half of both panels) against analytic solutions of the warm linear theory (bottom half of both panels), for both the two closures (top panel LEC, while bottom panel WARMC). The quantity shown is the particle density $n$, normalized w.r.t.~the initial plasma density $n_i$. We also plot the Mach cone structure, together with its inclination $\Theta$ for reference.}
    \label{fig:5}
\end{figure}
From a qualitative point of view, at this stage of the analysis there is no clear difference in behavior between the two fluid closures. A more quantitative characterization can be performed by analyzing these acoustic structures in the linear regime (i.e. for small values of the parameter $\tilde{Q}$), where it is possible to elaborate an analytical theory for both the two fluid models: in a nutshell, by using perturbation theory on respectively~\cref{eq:rel-euler} and~\cref{eq:warmc-fluid-eq1,eq:warmc-fluid-eq2,eq:warmc-fluid-eq3,eq:traces3,eq:traces4}, it is possible to derive a forced Klein-Gordon equation for the electron plasma density $n$, where the only difference between the fluid closures appear in the temperature dependent coefficients of the equation (see Appendix~\ref{app:warm-linear-theory} for details). By looking at this equation the acoustic nature of the dynamics becomes clear, and it is possible to derive some analytical results to compare with the numerical simulations. In Fig.~\ref{fig:5} we show the result of numerical simulations (top half of the panels) versus the analytical solutions (bottom half of the panel). There is good agreement between the two. By inspecting the two cited panels it is evident that the two fluid models, although exhibiting similar qualitative features, differ in the inclination $\Theta$ (plotted for reference in the figure) of the conical structure. The theory of super-sonic flows~\cite{landau-1987} tells us that this inclination is linked to the intensity of the velocity $c_s$ via
\begin{align}
c_s = c \sin (\Theta) \;,
\end{align}
which means that the wavefront of the perturbation propagates in the two models with different velocities. The warm linear theory provides in fact a dependency (look at Appendix~\ref{app:warm-linear-theory} for more analytical details) of $c_s$ from the initial temperature:
\begin{align}
    \label{eq:sound-speed1}
    \left(\frac{c_s}{c}\right)^2 &= \frac{5}{3} \mu_i    \quad \text{LEC}        \; , \\
    \label{eq:sound-speed2}
    \left(\frac{c_s}{c}\right)^2 &= 3 \mu_i              \quad \text{WARMC}      \; . 
\end{align}
One can then run a set of simulations with different initial temperatures $\mu_i$, and study the inclination of the conical envelope (from which we extract $c_s$) as a function of the temperature. In Fig.~\ref{fig:6} we show the results of this analysis, that again show good match between the simulations and expectations from theory. \\
As a final element of discussion we present the results of Fig.~\ref{fig:7}, where we consider the rest frame pressures ratio $P_{\parallel}/P_{\perp}$ for the WARMC model. This quantity is expected to be equal to one in the case of an isotropic pressure tensor, as it is the case in the LEC model, therefore in Fig.~\ref{fig:7} we consider the following quantity:
\begin{align}
    \frac{P_{\parallel}}{P_{\perp}}-1   \; .
\end{align}
The fluid rest frame pressures $P_\parallel$ and $P_\perp$ are obtained from the numerics by inverting~\cref{eq:double-boost} (the lab-frame components of the $\theta^{\mu\nu}$ tensor are obtained from the simulations). \\
We note thermal spread anisotropies are mostly relevant in the proximity of the bunch (located at $\zeta k_p=3.0$) where they become of the order of $10\%$, and become less and less important as one proceeds further away from the perturbation. To the best of our knowledge, this is the first time that such analysis is conducted on the WARMC model for a spatially resolved plasma (a study on a laser excited, 1D restricted plasma can be found in~\cite{khalilzadeh-2015}) and this preliminary results indicate that the LEC model, that is missing this pressure anisotropy feature by construction, could still be used to characterize plasma behavior in late-stage dynamics studies. We reserve further investigation, in particular studies on the dependency of this anisotropic feature on driving bunch parameters and initial temperatures, for later research.
%
\section{Outlook and perspectives} \label{sec:discussion}

In the development of a fluid model for the simulations of PWFA processes, a multitude of physical ingredients have to be taken into account to provide a realistic description of the process. In an earlier paper~\cite{parise-2022}, some of the authors started the exploration of lattice Boltzmann (LB)  schemes for the construction of fluid models for PWFA and considered the simplified case of cold fluid models. This paper represents a step forward, in that we have explored LB fluid schemes accounting for thermal effects~\cite{katsouleas-1988,rosenzweig-1988,schroeder-2005,esarey-2007,zgadzaj-2020,silva-2021,wang-2021,diederichs-2023,diederichs-2023-b}. The inclusion of thermal effects is a rather non trivial task, due to the theoretical complication represented by the choice of a proper closure to the set of equations~\cite{muscato-1993}. We have handled this problem by selecting two of the most popular closure models that have been discussed in the literature: the first one relies on the assumption of a local equilibrium (LEC)~\cite{toepfer-1971}, while the second one involves truncating the hierarchy of centered equations at an arbitrary order (WARMC)~\cite{schroeder-2010}. 
We have then shown how to successfully adapt the LB  schemes to both closures. If from one side the LEC is nominally not appropriate for a collisionless warm plasma, from the other side the WARMC is obtained under the assumption of an asymptotically small temperature. Any finite temperature, however small, can raise the question on what is the right closure scheme to obtain the correct fluid model for collisionless warm plasma dynamics. The preliminary comparisons shown in this work, although not presenting strong qualitative differences in the dynamics, give a clear indication that the selection of a closure scheme is pivotal for the quantitative assessment of PWFA experiments. To this aim, a one-to-one comparison between the predictions of fluid models and the PIC simulations (or a numerical solution of the Vlasov equations) could be helpful to shed lights on the matter. Work is in progress along this direction.\\
The next logical step in the development of the method is the inclusion of ions dynamics. Since plasma ions are way more massive then electrons, their dynamics happens on longer time scales than the ones examined in PWFA studies that target the early stage evolution of system, unless a strongly dense driving bunch is considered~\cite{rosenzweig-2005}. 
There is though a growing research interest for studies that inquire the late time evolution of the system where this kind of dynamics cannot be ignored anymore~\cite{darcy-2022,gorbunov-2003,gholizadeh-2011,khudiakov-2022}: unlocking the movement of ions brings in a new wealth of physical processes such as soliton dynamic~\cite{khudiakov-2022} and ion channels formation~\cite{zgadzaj-2020}. Furthermore, most of these studies invoke thermal effects to explain the acoustic motion of the ions~\cite{darcy-2022,khudiakov-2022,sahai-2017,silva-2021}, and it is then clear how the development of a method that would be able of handling both temperature and ions dynamics is of the uttermost importance. The inclusion of ion motion in the numerical scheme presented in this paper is theoretically trivial, and only brings in a bigger computational effort (more equations need to be integrated): the aforementioned characterization is therefore completely within the reach of the LB method.\\ 
We also would like to mention a couple of advancements that would make the method presented in this paper a more complete numerical tool for the simulation of wakefield acceleration processes: the first is represented by the possibility of including a laser source field as the plasma perturbating mechanism, as the LWFA has always been an alternative (w.r.t.~PWFA) route to wakefield acceleration~\cite{esarey-2009}; the second is the possibility of handling non-rigid particle bunches, as the tracking of the driving bunch properties is both an important element of diagnostic and also a feature to be kept under control in modern day PWFA experiments (see for example~\cite{pompili-2021}). Again, work in this direction is in progress. On the computational side, we would like to mention that the code developed for this work is amenable to a number of numerical optimizations (such for example porting on GPUs) and it is a firm intention of the authors to realize these improvements in future works.\\
As the most important design choice presented in this paper is the selection of the LB method as a fluid solver, a few comments are in place. In this paper we have shown that the LB method is well capable of recovering analytic results, and we would like to mention further advantages that could make LB a convenient and viable option in the context of PWFA simulations. Although PIC methods are able to model the dynamics at the particle level, and hence are able to model kinetic effects, they constantly have to keep at bay their inherent numerical noise~\cite{langdon-1979,birdsall-2018}. Therefore in some situations it would be preferable to dispose of alternative tools for prototyping, and to reserve more complete PIC simulations for later stages. Solvers based on a strict discretization of the Vlasov equations would not suffer of the noise problem and still be able to capture mesoscopic effects, but for any dimensionality more than 1D they would be too computationally demanding and hungry for memory resources. Fluid solvers (and the LB method among them) present then a viable option for the realization of quick PWFA simulations. They are coarse-grained and hence do not suffer of numerical noise, and still capture a good amount of the physical effects that are encountered in PWFA. 
\begin{figure}
    \centering
    \includegraphics[width=0.5\textwidth]{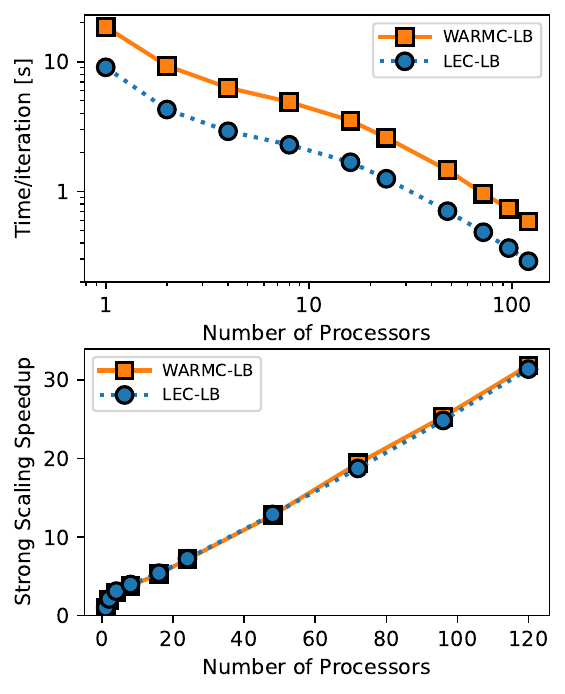}
    \caption{Top panel: execution time per iteration (in [s]) as a function of the number of processors for both LEC-LB (blue dotted) and WARMC-LB (orange solid). The scale is logarithmic. Bottom panel: corresponding speedups (data from the two closures are mostly overlapping). The simulation setup is the same as for results in Fig.~\ref{fig:4}-\ref{fig:5}.} 
    \label{fig:speedup}
\end{figure}
Furthermore, the LB method lends itself exceptionally well to parallelization on both CPUs and GPUs~\cite{krueger-2017}. In our implementation, we achieved parallelization on multi CPUs using the Message Passing Interface (MPI). This approach involves dividing the computational domain into multiple rectangular sub-domains, corresponding to the number of processors. Leveraging the local nature of computations (~\cref{eq:lbe}, r.h.s.), communications are solely required to exchange populations between neighboring processors during the ``streaming' process (~\cref{eq:lbe}, l.h.s.). By effectively disentangling ``compute'' and ``communicate'', this strategy greatly enhances the parallelization process~\cite{mazzeo-2008,bernaschi-2009,bernaschi-2010,succi-2019}.
Simulations in this study were conducted on an Intel Xeon E5-2695@2.40 GHz processor. A representative simulation (like those shown in Fig.~\ref{fig:4}-\ref{fig:5}) run on 96 processors requires approximately 182 minutes for the LEC-LB model and 368 minutes for the WARMC-LB model for $3 \cdot 10^4$ time steps. Memory requirements for such simulations are $\sim$ 1.6 GB for the LEC-LB model and $\sim$ 4.3 GB for the WARMC-LB model, although these numbers could be decreased by further optimizations of the code. As a final note, we would like to remark that LB takes roughly $50 \%$ of the compute time in the simulation, whereas the Maxwell solver for the electromagnetic fields takes $1\%$. This is to be expected as the amount of computation appearing in LB is significantly higher. Fig.~\ref{fig:speedup} presents the execution time per iteration and speedup data for varying numbers of processors (strong scaling).\\ 
Furthermore, an advantage of the LB method over Vlasov solvers is that the first adopts a smart pruning of the velocity space~\cite{krueger-2017,succi-2018}, thus improving the computational efficiency. It is possible to increase the number of discrete velocities $N_{pop}$ (see Sec.~\ref{sec:basic-lb}). However, this leads to a proportional increase in both the computational cost and memory requirements, making it a crucial factor to consider when choosing the LB stencil. Common practice is indeed to choose the minimum value $N_{pop}$ that ensures the recovery of the hydrodynamic properties~\cite{krueger-2017}. We hasten to remark, however, that at variance with usual hydrodynamic solvers, LB's theoretical formulation is strongly grounded in kinetic theory, and its fluid behavior is obtained via the smart discretization of the velocity space cited above. This is a remarkable feature that makes the method a strong candidate for the inclusion of kinetic effects into PWFA fluid solvers. In fact, recent studies~\cite{ambrus-2018, bazzanini-2021, ambrus-2022} in other research fields
show that LB is indeed capable of capturing behaviors beyond hydrodynamics just by increasing the number of discrete kinetic velocities. This could open novel perspectives for developing more refined numerical schemes for simulations of PWFA processes, with the obvious need of a precise comparison/benchmark against some reference PIC/Vlasov simulations. Further investigation on this is in progress. \\
We finally would like to mention that an alternative route to LB wakefield simulation could be represented by the Relativistic Lattice Boltzmann method~\cite{gabbana-2021}. This is an extension of LB hydrodynamic  schemes (like the ones exposed in Sec.~\ref{sec:basic-lb}) to the theory of special relativity, originally created for simulation in astrophysical~\cite{weih-2020} and condensed matter~\cite{gabbana-2018,simeoni-2023} contexts, and therefore constitutes a promising tool for the simulation of warm plasmas within the LEC assumption.
Adapting it to a PWFA framework is though more technical and less immediate w.r.t.~the {\it moment matching} LB used in this paper, therefore the authors reserve further development on this line for future works. 


\section*{Acknowledgements}

The authors gratefully acknowledge Fabio Bonaccorso for his technical support. This work was supported by the Italian Ministry of University and Research (MUR) under the FARE program (No. R2045J8XAW), project ``Smart-HEART''. MS gratefully acknowledges the support of the National Center for HPC, Big Data and Quantum Computing, Project CN\_00000013 - CUP E83C22003230001, Mission 4 Component 2 Investment 1.4, funded by the European Union - NextGenerationEU.


\section*{Author Declarations}

\subsection*{Conflict of Interests}
The authors declare that they have no conflict of interests.

\subsection*{Author Contributions}

\textbf{Daniele Simeoni}: Conceptualization (Lead), Formal Analysis (lead), Investigation (lead), Methodology (equal), Software (equal), Visualization (lead), Writing (lead). 
\textbf{Gianmarco Parise}: Conceptualization (supporting), Software (equal),  Formal Analysis (supporting), Writing (supporting). 
\textbf{Fabio Guglietta}: Conceptualization (supporting), Software (supporting), Writing (supporting).  
\textbf{Andrea Renato Rossi}: Conceptualization (supporting), Writing (supporting). 
\textbf{James Rosenzweig}: Conceptualization (supporting), Writing (supporting).  
\textbf{Alessandro Cianchi}: Conceptualization (supporting), Writing (supporting). 
\textbf{Mauro Sbragaglia}: Conceptualization (supporting), Methodology (equal), Writing (supporting).


\section*{Data Availability Statement}

The data that support the findings of this study are available from the corresponding author upon reasonable request.


\appendix

\section{Full expressions for the fluid advection equations}\label{app:full-eqs}

We report here the full form of the advection equations solved via the {\it moment matching} LB method in both the LEC and WARMC model. The equations are expressed in a 3D3V axisymmetric environment. In this context, we slightly modify the advection operator $\mathcal{D}_{\mybf{u}}$ defined in the main text to adopt its cylindrical counterpart:
\begin{align}
    \tilde{\mathcal{D}}_{\mybf{u}}(A) \equiv \partial_t A + \partial_r (A u_r) + \partial_z (A u_z) \; .
\end{align}
All cylindrical correction terms deriving from the divergence transformation are moved to the RHS of the equations, and are treated as source terms. The axisymmetry condition provides some useful simplifications:
\begin{enumerate}
    \item The azimuthal fluid velocity is zero: 
        \begin{align}
            0 = u_{\phi} \;.
        \end{align}
    \item Axisymmetric Maxwell equations lead to some electromagnetic field components to be zero: 
          \begin{align}
             0 = E_\phi=B_r=B_z \;.
          \end{align}
    \item In the WARMC model, thanks to the strategy explained in the main text (end of Sec.~\ref{sec:warmc-lb}) one       can derive rest frame quantities:
            \begin{align}
                0 &= \theta^{0\phi}=\theta^{r\phi}=\theta^{z\phi} \;, \\
                P_{\perp}       &= \theta^{\phi\phi}   \;, \\
                P_{\parallel}   &= -\frac{-c^2(\theta^{rr}+\theta^{zz})+\theta^{00}\mybf{u}^2+\theta^{\phi\phi}(c^2+u_r^2)}{c^2+u_z^2}  \; .
            \end{align}
\end{enumerate}

\subsection{LEC equations}

\begin{align}
    \tilde{\mathcal{D}}_{\mybf{u}} (n)      &= - \frac{u_r n}{r}                                    \;, \\
    \tilde{\mathcal{D}}_{\mybf{u}} (n A_r)  &= - \frac{u_r A_r}{r} 
    - n_i m_e c^2 \mu_i \partial_r \left(\frac{n}{\gamma n_i}\right)^{5/3} - e n (E_r - u_z B_{\phi}) \;, \\
    \tilde{\mathcal{D}}_{\mybf{u}} (n A_z)  &= - \frac{u_r A_z}{r}
    - n_i m_e c^2 \mu_i \partial_z \left(\frac{n}{\gamma n_i}\right)^{5/3} - e n (E_z + u_r B_{\phi}) \;, 
\end{align}

with 

\begin{align}
    \begin{pmatrix} 
        A_r \\ 
        A_z 
    \end{pmatrix}
    = \left[ 1 + \frac{5}{2} \mu_i \left( \frac{n}{\gamma n_i} \right)^{2/3} \right] n
    \begin{pmatrix} 
        p_r \\ 
        p_z 
    \end{pmatrix} \;.
\end{align}

\subsection{WARMC equations}

Define $\lambda = n/h$:

\begin{align}
    \tilde{\mathcal{D}}_{\mybf{u}}(n)    &= -\frac{n u_r}{r}                                                \;, \\
    \tilde{\mathcal{D}}_{\mybf{u}} ( n \lambda c ) &= 
                                                    - \frac{n \lambda c u_r}{r}
                                                    - \frac{1}{c} \partial_t \theta^{00} 
                                                    - \partial_r \theta^{0r} 
                                                    - \partial_z \theta^{0z}
                                                    - \frac{\theta^{0r}}{r}                             
                                                    - \frac{en}{c} (E_r u_r + E_z u_z)                      \;, \\
    \tilde{\mathcal{D}}_{\mybf{u}} ( n \lambda  u_r ) &=
                                                - \frac{n \lambda u_r^2}{r}  
                                                - \frac{1}{c} \partial_t \theta^{0r} 
                                                - \partial_r \theta^{rr} 
                                                - \partial_z \theta^{rz}
                                                - \frac{\theta^{rr}-\theta^{\phi\phi}}{r}  
                                                - e n (E_r - u_z B_\phi )                                    \;, \\
    \tilde{\mathcal{D}}_{\mybf{u}} ( n \lambda  u_z ) &= 
                                                - \frac{n \lambda u_z u_r }{r}
                                                - \frac{1}{c} \partial_t \theta^{0z}  
                                                - \partial_r \theta^{rz} 
                                                - \partial_z \theta^{zz} 
                                                - \frac{\theta^{rz}}{r}
                                                - e n (E_z + u_r B_\phi)                                     \;, \\
    \tilde{\mathcal{D}}_{\mybf{u}} ( \lambda \theta^{rr} ) &= 
                                        - \frac{ \lambda \theta^{rr} u_r}{r}
                                        - 2 \frac{\theta^{0r}}{c} \partial_t \left( \lambda u_r  \right)
                                        - 2 \theta^{rr} \partial_r \left( \lambda u_r  \right)
                                        - 2 \theta^{rz} \partial_z \left( \lambda u_r  \right)
                                        - 2 \frac{e}{c} (E_r \theta^{0r} - c B_\phi \theta^{rz} )            \;, \\
    \tilde{\mathcal{D}}_{\mybf{u}} ( \lambda \theta^{\phi\phi}) ) &= 
                                          - 3 \frac{\lambda \theta^{\phi\phi} u_r}{r}                        \;, \\
    \tilde{\mathcal{D}}_{\mybf{u}} (\lambda \theta^{zz}) &= 
                                          - \frac{ \lambda \theta^{zz} u_r}{r}
                                          - 2 \frac{\theta^{0z}}{c} \partial_t \left( \lambda  u_z  \right)
                                          - 2 \theta^{rz} \partial_r \left( \lambda  u_z  \right)
                                          - 2 \theta^{zz} \partial_z \left( \lambda  u_z  \right)
                                          - 2 \frac{e}{c} (E_z \theta^{0z} + c B_\phi \theta^{rz} )          \;, \\
    \tilde{\mathcal{D}}_{\mybf{u}} (\lambda \theta^{rz}) &= 
                                          - \frac{ \lambda \theta^{rz} u_r}{r}
                                          - \frac{\theta^{0z}}{c} \partial_t \left( \lambda  u_r  \right)
                                          - \frac{\theta^{0r}}{c} \partial_t \left( \lambda  u_z  \right)
                                          - \theta^{rz} \partial_r \left( \lambda  u_r  \right)
                                          - \theta^{rr} \partial_r \left( \lambda  u_z  \right)
                                          - \theta^{zz} \partial_z \left( \lambda  u_r  \right)
                                          - \theta^{rz} \partial_z \left( \lambda  u_z  \right)         \notag \\
                                         &- \frac{e}{c} [E_r \theta^{0z} + E_z \theta^{0r} 
                                          + c B_\phi (\theta^{rr} - \theta^{zz}) ]                              \;.
\end{align}
The mass-shell conditions~\cref{eq:traces3,eq:traces4} provide the remaining non-zero components of the tensor $\theta^{\mu\nu}$:
\begin{align}
    \theta^{0r} &= \frac{u_r}{c} \theta^{rr} + \frac{u_z}{c} \theta^{rz}    \;, \\
    \theta^{0z} &= \frac{u_r}{c} \theta^{rz} + \frac{u_z}{c} \theta^{zz}    \;, \\
    \theta^{00} &= \frac{u_r}{c} \theta^{0r} + \frac{u_z}{c} \theta^{0z}    \;.
\end{align}

\section{Warm linear theory}\label{app:warm-linear-theory}

The fluid equations, respectively~\cref{eq:rel-euler} and~\cref{eq:warmc-fluid-eq1,eq:warmc-fluid-eq2,eq:warmc-fluid-eq3,eq:traces3,eq:traces4} for the two closures, can be linearly perturbed w.r.t.~the initial rest state, when coupled with Maxwell equations~\cref{eq:maxwell-homogeneus,eq:maxwell-inhomogeneus}. One then obtains a forced Klein-Gordon equation for the density perturbation $n_1=n-n_i$:
\begin{align}\label{eq:fhow}
    \partial_t^2 n_1 - c_s^2 \nabla^2 n_1 = - a_e^2 n_1 - a_b^2 n_b ~.
\end{align}
Where the coefficients $c_s$, $a_e$ and $a_b$ depend on the initial temperature in a way that's dictated by the selected closure scheme. At order $O(\mu_i)$, one has:
\begin{align}
     \textbf{LEC}&  \quad
     \begin{cases}
          \left( \frac{c_s}{c}          \right)^2                     &= \frac{5}{3} \mu_i      \\ 
          \left( \frac{a_e}{\omega_p}   \right)^2                     &= 1 - \frac{5}{2} \mu_i  \\
          a_b^2                     &= a_e^2
     \end{cases} \; ,
\end{align}
\begin{align}
     \textbf{WARMC}& \quad 
     \begin{cases}
          \left( \frac{c_s}{c}          \right)^2 &= 3 \mu_i               \\
          \left( \frac{a_e}{\omega_p}   \right)^2 &= 1 - \frac{5}{2} \mu_i \\
          \left( \frac{a_b}{\omega_p}   \right)^2 &= 1 - \frac{1}{2} \mu_i
     \end{cases} \; .
\end{align}
This result, aside from giving an explicit dependency of the acoustic velocity $c_s$ w.r.t.~temperature (result~\cref{eq:sound-speed1,eq:sound-speed2}), tells us that the behavior of the two fluid schemes is qualitatively the same. One can further inspect~\cref{eq:fhow} to determine a dispersion relation for the Langmuir wave. Assuming to be far away from the driving bunch (so that $n_b$ can be ignored), one derives the following by recurring to Fourier transformation $(\partial_t, \mybf{\nabla}) \rightarrow (-i \omega, i \mybf{k})$:
\begin{align}
    \omega^2 = a_e^2 + c_s^2 \mybf{k}^2  \; ,
\end{align}
which reduces to the result~\cref{eq:disp-rel1,eq:disp-rel2} when considered at order $O(\mu_i)$. Last thing to discuss is the strategy to the solution of~\cref{eq:fhow}. After performing the traveling wave ansatz, which states that every field is dependent on $z$ and $t$ only through the co-moving coordinate $\zeta=z-ct$, we obtain the following equation:
\begin{align}\label{eq:fhow-comoving}
(c^2 -c_s^2) \partial_\zeta^2 n_1 - c_s^2 \nabla_{\perp}^2 n_1 = - a_e^2 n_1 - a_b^2 n_b \; .
\end{align}
At this point, we introduce the \textit{Hankel Tranformation of order 0}~\cite{offord-1935}:
\begin{align}
     H_0 \left[ f(r) \right] = \hat{f}(w) = \int_{0}^{+\infty} f(r) r J_0 (w r) dr  \; ,
\end{align}
where $J_0 (wr)$ is the \textit{Bessel Function of first kind and order 0}. This transform behaves nicely in the presence of transversal laplacians of radial functions $(\nabla_{\perp})\rightarrow (i w)$. Therefore, if one Hankel transforms~\cref{eq:fhow-comoving}, a Forced Harmonic oscillator in the variable $\zeta$ is obtained:
\begin{align}
     \partial_\zeta^2 \hat{n}_1(w,\zeta) = 
     &- \left( \frac{a_e^2 + w^2 c_s^2}{c^2-c_s^2} \right) \hat{n}_1(w,\zeta) \notag\\
     &- \left( \frac{a_b^2}{c^2-c_s^2} \right) \hat{n}_b(w,\zeta)   \; ,
\end{align}
which can be solved via the Gauss-Green method. The final anti-transformed solution, to be numerically integrated, reads therefore as:
\begin{align}
    n_1(r, \zeta) = - \frac{a_b^2}{\sqrt{c^2-c_s^2}} \int_{-\infty}^{\zeta} \int_{0}^{+\infty} \frac{\hat{n}_b(w,\zeta')J_0(w r)}{\sqrt{a_e^2+w^2c_s^2}} \sin\left[\sqrt{\frac{a_e^2 + w^2 c_s^2}{c^2-c_s^2}}(\zeta-\zeta')\right] w dw d\zeta' \;.
\end{align}

\printbibliography

\vfill
\begin{center}
    \tiny{This article may be downloaded for personal use only. Any other use requires prior permission of the author and AIP Publishing. 
    \\ This article appeared in~\cite{simeoni-2024} and may be found at \url{https://doi.org/10.1063/5.0175910}}
\end{center}

\end{document}